\documentclass[amsmath,amssymb,12pt,nofootinbib]{article}
\pdfoutput=1
\usepackage{jheppub}

\usepackage{hyperref}
\usepackage{amsmath}
\usepackage{graphicx}
\usepackage{xspace}
\usepackage{color}
\usepackage{overpic}
\usepackage{url}
\usepackage[utf8]{inputenc}
\usepackage{epstopdf}
\usepackage{tikz}
\usepackage{caption}
\usepackage{makecell}
\usepackage{bm}
\usepackage{revsymb}
\usepackage{slashed}
\usepackage{float}
\usepackage{tabularx}
\usepackage{subfig}

\usepackage{fixltx2e} % For \textsubscript

\makeatletter
\DeclareRobustCommand*\textsubsuperscript[2]{%
	\@textsubsuperscript{\selectfont#1}{\selectfont#2}}
\def\@textsubsuperscript#1#2{%
	{\m@th\ensuremath{_{\mbox{\fontsize\sf@size\z@#1}}
			^{\mbox{\fontsize\sf@size\z@#2}}}}}
\makeatother

\newcommand{\bea}{\begin{eqnarray}}
\newcommand{\eea}{\end{eqnarray}}

\newcommand{\eq}[1]{Eq.~\eqref{#1}}

\definecolor{point1}{rgb}{1,0,0}
\definecolor{point2}{rgb}{0.5,0,0.5}
\definecolor{point3}{rgb}{1,0.75,0}
\definecolor{point4}{rgb}{0,0.66,0}

\begin{document}
	\preprint{PSI-PR-20-08 , ZU-TH 20/20}
	
	\title{Leptoquarks in Oblique Corrections and Higgs Signal Strength: Status and Prospects}
	
	\author[a,b]{Andreas Crivellin}
	\affiliation[a]{Physik-Institut, Universit\"at Z\"urich,
		Winterthurerstrasse 190, CH-8057 Z\"urich, Switzerland}
	\affiliation[b]{Paul Scherrer Institut, CH--5232 Villigen PSI, Switzerland}
	
	\author[a,b]{Dario M\"uller}
	\affiliation[a]{Physik-Institut, Universit\"at Z\"urich,
		Winterthurerstrasse 190, CH-8057 Z\"urich, Switzerland}
	
	\author[c]{Francesco Saturnino}
	\affiliation[c]{Albert Einstein Center for Fundamental Physics, Institute
		for Theoretical Physics,\\ University of Bern, CH-3012 Bern,
		Switzerland}
	
	\emailAdd{andreas.crivellin@cern.ch}
	\emailAdd{dario.mueller@psi.ch}
	\emailAdd{saturnino@itp.unibe.ch}
	
\abstract{Leptoquarks (LQs) are predicted within Grand Unified Theories and are well motivated by the current flavor anomalies. In this article we investigate the impact of scalar LQs on Higgs decays and oblique corrections as complementary observables in the search for them. Taking into account all five LQ representations under the Standard Model gauge group and including the most general mixing among them, we calculate the effects in $h\to\gamma\gamma$, $h\to gg$, $h \to Z \gamma$ and the Peskin-Takeuchi parameters $S$, $T$ and $U$. We find that these observables depend on the same Lagrangian parameters, leading to interesting correlations among them. While the current experimental bounds only yield weak constraints on the model, these correlations can be used to distinguish different LQ representations at future colliders (ILC, CLIC, FCC-ee and FCC-hh), whose discovery potential we are going to discuss.}
	

\maketitle
	
\section{Introduction}
	
Leptoquarks (LQs) are particles which have a specific interaction vertex, connecting a lepton with a quark. They are predicted in Grand Unified Theories~\cite{Pati:1974yy,Georgi:1974sy,Georgi:1974yf,Fritzsch:1974nn} and were systematically classified in Ref.~\cite{Buchmuller:1986zs} into ten possible representations under the Standard Model (SM) gauge group (five scalar and five vector particles). In recent years, LQs experienced a renaissance due to the emergence of the flavor anomalies. In short, hints for new physics (NP) in $R(D^{(*)})$~\cite{Lees:2012xj,Lees:2013uzd,Aaij:2015yra,Aaij:2017deq,Aaij:2017uff,Abdesselam:2019dgh}, $b\to s\ell^{+}\ell^{-}$~\cite{CMS:2014xfa,Aaij:2015oid,Abdesselam:2016llu,Aaij:2017vbb,Aaij:2019wad,Aaij:2020nrf} and $a_\mu$~\cite{Bennett:2006fi} emerged, with a significance of $>3\,\sigma$~\cite{Amhis:2016xyh,Murgui:2019czp,Shi:2019gxi,Blanke:2019qrx,Kumbhakar:2019avh}, $>5\sigma$~\cite{Capdevila:2017bsm, Altmannshofer:2017yso,Alguero:2019ptt,Alok:2019ufo,Ciuchini:2019usw,Aebischer:2019mlg, Arbey:2019duh,Kumar:2019nfv} and $>3\,\sigma$~\cite{Aoyama:2020ynm}, respectively. It has been shown that LQs can explain $b\to s\ell^+\ell^-$ data~\cite{Alonso:2015sja, Calibbi:2015kma, Hiller:2016kry, Bhattacharya:2016mcc, Buttazzo:2017ixm, Barbieri:2015yvd, Barbieri:2016las, Calibbi:2017qbu, Crivellin:2017dsk, Bordone:2018nbg, Kumar:2018kmr, Crivellin:2018yvo, Crivellin:2019szf, Cornella:2019hct, Bordone:2019uzc, Bernigaud:2019bfy,Aebischer:2018acj,Fuentes-Martin:2019ign,Fajfer:2015ycq,  Blanke:2018sro,deMedeirosVarzielas:2019lgb,Varzielas:2015iva,Crivellin:2019dwb,Saad:2020ihm,Saad:2020ucl}, $R(D^{(*)})$~\cite{Alonso:2015sja, Calibbi:2015kma, Fajfer:2015ycq, Bhattacharya:2016mcc, Buttazzo:2017ixm, Barbieri:2015yvd, Barbieri:2016las, Calibbi:2017qbu, Bordone:2017bld, Bordone:2018nbg, Kumar:2018kmr, Biswas:2018snp, Crivellin:2018yvo, Blanke:2018sro, Heeck:2018ntp,deMedeirosVarzielas:2019lgb, Cornella:2019hct, Bordone:2019uzc,Sahoo:2015wya, Chen:2016dip, Dey:2017ede, Becirevic:2017jtw, Chauhan:2017ndd, Becirevic:2018afm, Popov:2019tyc,Fajfer:2012jt, Deshpande:2012rr, Freytsis:2015qca, Bauer:2015knc, Li:2016vvp, Zhu:2016xdg, Popov:2016fzr, Deshpand:2016cpw, Becirevic:2016oho, Cai:2017wry, Altmannshofer:2017poe, Kamali:2018fhr, Azatov:2018knx, Wei:2018vmk, Angelescu:2018tyl, Kim:2018oih, Crivellin:2019qnh, Yan:2019hpm,Crivellin:2017zlb, Marzocca:2018wcf, Bigaran:2019bqv,Crivellin:2019dwb,Saad:2020ihm,Dev:2020qet,Saad:2020ucl,Altmannshofer:2020axr} and/or $a_\mu$~\cite{Bauer:2015knc,Djouadi:1989md, Chakraverty:2001yg,Cheung:2001ip,Popov:2016fzr,Chen:2016dip,Biggio:2016wyy,Davidson:1993qk,Couture:1995he,Mahanta:2001yc,Queiroz:2014pra,ColuccioLeskow:2016dox,Chen:2017hir,Das:2016vkr,Crivellin:2017zlb, Cai:2017wry, Crivellin:2018qmi,Kowalska:2018ulj,Mandal:2019gff,Dorsner:2019itg,Crivellin:2019dwb,DelleRose:2020qak,Saad:2020ihm,Bigaran:2020jil}.
\smallskip

This strong motivation for LQs makes it also interesting to search for their signatures in other observables. Complementary to direct LHC searches~\cite{Kramer:1997hh,Kramer:2004df,Faroughy:2016osc, Greljo:2017vvb, Dorsner:2017ufx, Cerri:2018ypt, Bandyopadhyay:2018syt, Hiller:2018wbv, Faber:2018afz, Schmaltz:2018nls,Chandak:2019iwj,Allanach:2019zfr, Buonocore:2020erb,Borschensky:2020hot}, oblique electroweak (EW) parameters ($S$ and $T$ parameters~\cite{Peskin:1991sw,Altarelli:1990zd}) and the corrections to (effective on-shell) couplings of the SM Higgs to photons ($h\gamma\gamma$), $Z$ and photon ($h Z\gamma$) and gluons ($hgg$) allow to test LQ interactions with the Higgs, independently of the LQ couplings to fermions. In this context, LQs were briefly discussed in Ref.~\cite{Dorsner:2016wpm} based on analogous MSSM calculations~\cite{Djouadi:2005gj,Muhlleitner:2006wx,Bonciani:2007ex}, simplified model analysis~\cite{Carena:2012xa,Chang:2012ta,Gori:2013mia,Chen:2013vi}, vacuum stability \cite{Bandyopadhyay:2016oif}, LQ production at hadron colliders~\cite{Agrawal:1999bk} and Higgs pair production~\cite{Enkhbat:2013oba}. In addition, Ref.~\cite{Bhaskar:2020kdr} recently studied LQs in Higgs production and Ref.~\cite{Zhang:2019jwp} considered $h\to\gamma\gamma$, while Ref.~\cite{Gherardi:2020det} performed the matching in the singlet-triplet model \cite{Crivellin:2017zlb}. However, none of these analyses considered more than a single LQ representation at a time. The situation is similar concerning the $S$ and $T$ parameter. This was also briefly discussed in Ref.~\cite{Dorsner:2016wpm}, based on simplified model calculations~\cite{Froggatt:1991qw} and an analysis discussing only the $SU(2)_L$ doublet LQs~\cite{Keith:1997fv}. Most importantly, the unavoidable correlations between Higgs couplings to gauge bosons and the oblique parameters were not considered so far. Importantly, these observables can be measured much more precisely at future colliders such as the ILC~\cite{Behnke:2013lya}, CLIC~\cite{Aicheler:2012bya}, and the FCC~\cite{Abada:2019lih,Abada:2019zxq}. Therefore, it is interesting to examine their estimated constraining power and discovery potential. 
\smallskip
	
In this article we will calculate the one-loop effects of LQs in oblique corrections, $h\gamma\gamma$, $h Z\gamma$ and $hgg$, taking into account all five scalar LQ representations and the complete set of their interactions with the Higgs. In the next section we will define our setup and conventions before we turn to the calculation of the $S$ and $T$ parameters in Sec.~\ref{sec:Oblique} and to $h\gamma\gamma$, $h Z\gamma$ and $hgg$ in Sec.~\ref{sec:Higgs}. We then perform our phenomenological analysis, examining the current status and future prospects for these observables in Sec.~\ref{sec:Pheno}, before we conclude in Sec.~\ref{sec:Conclusions}. An appendix provides useful analytic (perturbative) expressions for LQ couplings and results for the loop functions.

\section{Setup and Conventions}

There are ten possible representations of LQs under the SM gauge group~\cite{Buchmuller:1986zs}. While for vector LQs a Higgs mechanism is necessary to render the model renormalizable, scalar LQs can simply be added to the SM. Since we are interested in loop effects in this work, we will focus on the latter ones in the following.
\smallskip

The five different scalar LQs transform under the SM gauge group
\begin{align}
\mathcal{G}_{\text{SM}}=SU(3)_{c}\times SU(2)_L\times U(1)_Y
\end{align}
as given in Table~\ref{eq:SLQ_interaction}. 
\begin{table}
	\centering
	\renewcommand{\arraystretch}{1.8}
	\begin{tabular}{c|c|c|c|c|c}
		& $\Phi_1$ & $\tilde{\Phi}_1$ & $\Phi_{2}$ & $\tilde{\Phi}_2$ & $\Phi_{3}$\\
		\hline
		$\mathcal{G}_{\text{SM}}$  
		& $\bigg(3,1,-\dfrac{2}{3}\bigg)$ 
		& $\bigg(3,1,-\dfrac{8}{3}\bigg)$ 
		& {$\bigg(3,2,\dfrac{7}{3}\bigg)$} 
		& $\bigg(3,2,\dfrac{1}{3}\bigg)$ 
		& $\bigg(3,3,-\dfrac{2}{3}\bigg)$
	\end{tabular}
	\caption{LQ representations under the SM gauge group.}
	\label{eq:SLQ_interaction}
\end{table}

We defined the hypercharge $Y$ such that the electromagnetic charge is given by
\begin{align}
Q=\frac{1}{2}Y+T_{3}\,,
\end{align}
with $T_{3}$ representing the third component of weak isospin, e.g. $\pm 1/2$ for $SU(2)_L$ doublets and $1,0,-1$ for the $SU(2)_L$ triplet. Therefore, we have the following eigenstates with respect to the electric charge
\begin{align}
	\begin{aligned}
	&\Phi_{1}\equiv\Phi_{1}^{-1/3} \,,&&
	\tilde{\Phi}_{1}\equiv \tilde{\Phi}_{1}^{-4/3}\,,\\
	&\Phi_{2}\equiv\begin{pmatrix} \Phi_{2}^{5/3}\\\Phi_{2}^{2/3}\end{pmatrix}\,,&& 
	\tilde{\Phi}_{2}\equiv \begin{pmatrix}\tilde{\Phi}_{2}^{2/3}\\\tilde{\Phi}_{2}^{-1/3}\end{pmatrix}\,,&&
	\tau\cdot\Phi_{3}\equiv
	\begin{pmatrix}\Phi_{3}^{-1/3}& \sqrt{2}\Phi_{3}^{2/3}\\ \sqrt{2}\Phi_{3}^{-4/3}&-\Phi_{3}^{-1/3}\end{pmatrix}\,,
	\end{aligned}
\end{align}
obtained from the five representations. Note that the upper index refers to the electric charge and the lower one to the $SU(2)_{L}$ representation from which the field originates.
\smallskip

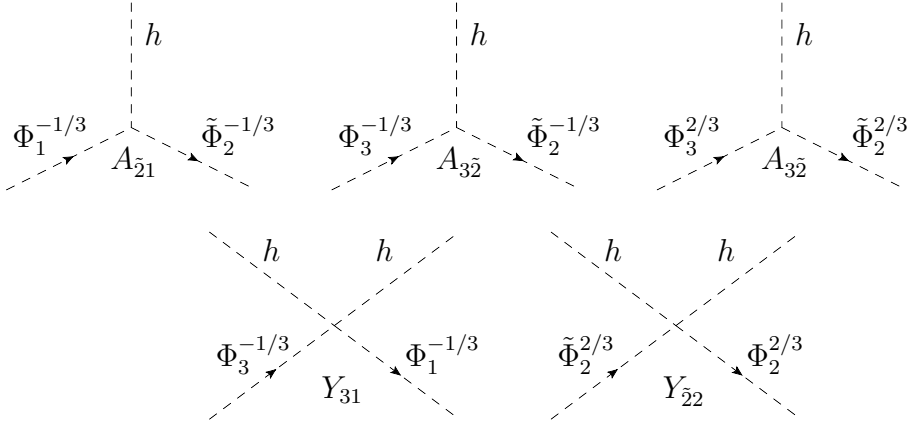
\begin{figure}
	\centering
	\begin{overpic}[scale=.47,,tics=10]
		{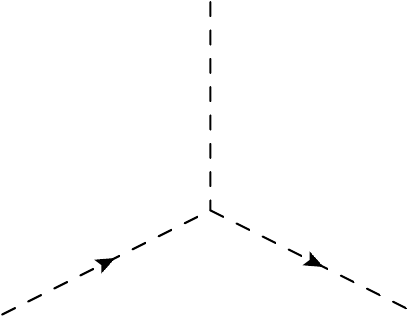}
		\put(3,18){$\Phi_{1}^{-1/3}$}
		\put(80,18){$\tilde{\Phi}_{2}^{-1/3}$}
		\put(43,9){$A_{\tilde 21}$}
		\put(57,60){$h$}
	\end{overpic}
\hspace{0.8cm}
	\begin{overpic}[scale=.47,,tics=10]
		{LQ_LQ_higgs}
		\put(3,18){$\Phi_{3}^{-1/3}$}
		\put(80,18){$\tilde{\Phi}_{2}^{-1/3}$}
		\put(43,9){$A_{3\tilde 2}$}
		\put(57,60){$h$}
	\end{overpic}
\hspace{0.8cm}
	\begin{overpic}[scale=.47,,tics=10]
		{LQ_LQ_higgs}
		\put(3,18){$\Phi_{3}^{2/3}$}
		\put(80,18){$\tilde{\Phi}_{2}^{2/3}$}
		\put(43,9){$A_{3\tilde 2}$}
		\put(57,60){$h$}
	\end{overpic}
\vspace{0.5cm}
\newline
	\begin{overpic}[scale=.47,,tics=10]
		{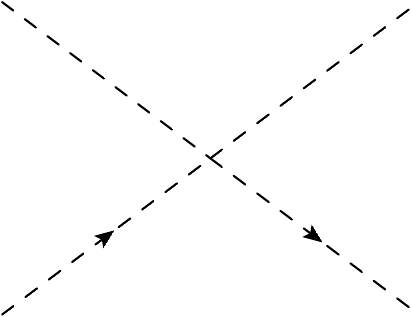}
		\put(3,22){$\Phi_{3}^{-1/3}$}
		\put(80,22){$\Phi_{1}^{-1/3}$}
		\put(46,9){$Y_{31}$}
		\put(68,65){$h$}
		\put(22,65){$h$}
	\end{overpic}
\hspace{1cm}
	\begin{overpic}[scale=.47,,tics=10]
		{LQ_LQ_higgs_higgs}
		\put(3,22){$\tilde{\Phi}_{2}^{2/3}$}
		\put(80,22){$\Phi_{2}^{2/3}$}
		\put(46,9){$Y_{\tilde 22}$}
		\put(68,65){$h$}
		\put(22,65){$h$}
	\end{overpic}
	\caption{Feynman diagrams depicting LQ-Higgs interactions. Here the physical Higgs $h$ can be replaced by its vev, leading to mixing among the LQs.}
	\label{fig:diagramm_HLQLQ_int}
\end{figure}

In addition to the gauge interactions of the LQs, determined by the respective representation under the SM gauge group, LQs can couple to the SM Higgs doublet $H$ (with hypercharge +1) via the Lagrangian~\cite{Hirsch:1996qy}\footnote{$Y_{\tilde{2}\tilde{2}}$ and $Y_{{2}{2}}$ were studied in Ref.~\cite{Keith:1997fv} while the $Y_{33}$ term was considered in Ref.~\cite{Gherardi:2020qhc}.}
\begin{align}
\begin{aligned}
	\mathcal{L}_{H\Phi}&=-A_{\tilde 2 1}\big( \tilde{\Phi}_2^{\dagger} H\big)\Phi_1+A_{3 \tilde 2}\big(\tilde{\Phi}_2^{\dagger}\big(\tau\cdot\Phi_{3}\big)H\big)+Y_{\tilde 22}\big(\Phi_{2}^{\dagger}H\big)\big(Hi\tau_{2}\tilde{\Phi}_{2}\big)\\
	&+Y_{3 \tilde 1}\big(H i\tau_{2} \left(\tau\cdot\Phi_{3}\right)^\dagger H\big)\tilde{\Phi}_1+Y_{3 1} \big(H^{\dagger}\left(\tau\cdot\Phi_{3} \right)H \big)\Phi_{1}^{\dagger}+\text{h.c.}\\
	&-Y_{22}\big(Hi\tau_{2}\Phi_{2}\big)\big(Hi\tau_{2}\Phi_{2}\big)^{\dagger}-Y_{\tilde{2}\tilde{2}}\big(Hi\tau_{2}\tilde{\Phi}_{2}\big)\big(Hi\tau_{2}\tilde{\Phi}_{2}\big)^{\dagger}\\
	&-iY_{33}\varepsilon_{IJK}H^{\dagger}\tau_{I}H\Phi_{3,J}^{\dagger}\Phi_{3,K}\\
	&-\sum_{k=1}^{3}\big(m_{k}^2+Y_{k} H^{\dagger}H\big)\Phi_{k}^{\dagger}\Phi_{k}-\sum_{k=1}^{2}\big(\tilde{m}_{k}^2+Y_{\tilde k} H^{\dagger}H\big)\tilde{\Phi}_{k}^{\dagger}\tilde{\Phi}_{k}\,.
	\label{eq:Higgs_LQ_lagrangian}
\end{aligned}
\end{align}
Here $m_{\Phi}^2$ represent the usual (bare) mass terms of the LQs, present without EW symmetry breaking and $\varepsilon_{IJK}$ is the three-dimensional Levi-Civita tensor with $\varepsilon_{123}=1$. Note that $A_{\tilde 21}$ and $A_{3\tilde 2}$ have mass dimension one, while the $Y$ couplings are dimensionless. The LQ-Higgs interactions lead to additional contributions to the mass matrices. The mixing among them is depicted in Figure~\ref{fig:diagramm_HLQLQ_int}.
\smallskip

Once the Higgs acquires a vacuum expectation value (vev) with $v\approx 174$ GeV, this generates the following mass matrices in the interaction basis
\begin{align}
\begin{aligned}
\mathcal{M}^{-1/3}&=
\begin{pmatrix}
m_{1}^2+v^{2}Y_{1}& vA_{\tilde 21}^*&v^2Y_{31}\\
vA_{\tilde 21}&\tilde{m}_{2}^2+v^{2}Y_{\tilde{2}}&vA_{3\tilde 2}\\
v^2Y_{31}^*&vA_{3\tilde 2}^*&m_{3}^2+v^{2}Y_{3}
\end{pmatrix}\,,\\
\mathcal{M}^{2/3}&=
\begin{pmatrix}
m_{2}^2+v^{2}Y_{2}&v^2Y_{\tilde 22}&0\\
v^2Y_{\tilde 22}^*&\tilde{m}_{2}^2+v^{2}\big(Y_{\tilde{2}\tilde{2}}+Y_{\tilde{2}}\big)& -\sqrt{2}vA_{3\tilde 2}\\
0& -\sqrt{2}vA_{3\tilde 2}^*&m_{3}^2+v^{2}\big(Y_{3}+Y_{33}\big)
\end{pmatrix}\,,
\\
\mathcal{M}^{-4/3}&=
\begin{pmatrix}
\tilde{m}_{1}^2+v^{2}Y_{\tilde{1}}&\sqrt{2}v^2Y_{3 \tilde 1}^*\\
\sqrt{2}v^2Y_{3 \tilde 1}&m_{3}^2+v^{2}(Y_{3}-Y_{33})
\end{pmatrix}\,,
\\
\mathcal{M}^{5/3}&=m_{2}^2+v^{2}\big(Y_{22}+Y_{2}\big)\,,
\end{aligned}
\label{eq:LQ_mixing_matrices}
\end{align}
such that
\begin{align}
-\,\Phi^{\dagger}_{Q}\mathcal{M}^{Q}\Phi_{Q}\subset \mathcal{L}_{H\Phi}\,.
\end{align}
This now parametrizes the mass terms in the Lagrangian, where $Q$ is the electric charge and we defined
\begin{align}
	\Phi_{-1/3}\equiv
	\begin{pmatrix}
	\Phi_{1}^{-1/3}\\ \tilde{\Phi}_2^{-1/3}\\ \Phi_{3}^{-1/3}
	\end{pmatrix}&&
	\Phi_{2/3}\equiv
	\begin{pmatrix}
	\Phi_{2}^{2/3}\\ \tilde{\Phi}_{2}^{2/3}\\ \Phi_3^{2/3}
	\end{pmatrix}&&
	\Phi_{-4/3}\equiv
	\begin{pmatrix}
	\tilde{\Phi}_{1}^{-4/3}\\ \Phi_{3}^{-4/3}
	\end{pmatrix}&&
	\Phi_{5/3}\equiv\Phi_{2}^{5/3}\,.
	\label{EWeigenstates}
\end{align}

In order to arrive at the physical basis we need to diagonalize the mass matrices in~\eq{eq:LQ_mixing_matrices}. This can be achieved via 
\begin{align}
\hat{\mathcal{M}}^{Q}=W^{Q}\mathcal{M}^{Q}W^{Q\dagger}
\end{align}
with unitary matrices $W^{Q}$. Thus, the interaction eigenstates in~\eq{EWeigenstates} are rotated as
\begin{align}
W^{Q}\Phi_{Q}\equiv\hat{\Phi}^{Q}
\end{align}
to arrive at the mass eigenstates. The matrices $W^{Q}$ for $Q=-1/3$ and $Q=2/3$ too lengthy to be given analytically in full generality, but can of course be computed numerically. However, in order to obtain the explicit dependence on the Lagrangian parameters $A$ and $Y$, we diagonalize the mass matrices perturbatively up to $\mathcal{O}(v^2)$, which then yields the following expressions
\begin{align}
W^{-1/3}&\approx\begin{pmatrix}
1\!-\!\frac{v^2|A_{\tilde 21}|^2}{2(m_{1}^2-\tilde{m}_2^2)^2}& \frac{v A_{\tilde 21}^{*}}{m_{1}^2-\tilde{m}_{2}^2} & \frac{v^2(Y_{31}(m_{1}^2-\tilde{m}_2^2)+A_{\tilde 21}^{*}A_{3\tilde 2})}{(m_{1}^2-m_{3}^2)(m_{1}^2-\tilde{m}_{2}^2)}\\
\frac{-v A_{\tilde 21}}{m_{1}^2-\tilde{m}_{2}^2} & 1\!-\!\frac{v^2}{2}\Big(\!\frac{|A_{\tilde 21}|^2}{(m_{1}^2-\tilde{m}_{2}^2)^2}\!+\!\frac{|A_{3\tilde 2}|^2}{(m_{3}^2-\tilde{m}_{2}^2)^2}\!\Big)&
\frac{-v A_{3\tilde 2}}{m_{3}^2-\tilde{m}_{2}^2}\\
\frac{-v^2(Y_{31}^{*}(m_{3}^2-\tilde{m}_2^2)+A_{\tilde 21}A_{3\tilde 2}^{*})}{(m_{1}^2-m_{3}^2)(m_{3}^2-\tilde{m}_{2}^2)}&\frac{v A_{3\tilde 2}^{*}}{m_3^2-\tilde{m}_{2}^2}& 1\!-\!\frac{v^2 |A_{3\tilde 2}|^2}{2(m_{3}^2-\tilde{m}_{2}^2)^2}
\end{pmatrix}\,,\nonumber\\
W^{2/3}&\approx\begin{pmatrix}
1& \frac{v^2 Y_{\tilde 22}}{m_{2}^2-\tilde{m}_{2}^2} & 0\\
\frac{-v^2 Y_{\tilde 22}^{*}}{m_{2}^2-\tilde{m}_{2}^2} & 1\!-\!\frac{v^{2}|A_{3\tilde 2}|^2}{(m_{3}^2-\tilde{m}_{2}^2)^2} & \frac{-\sqrt{2}vA_{3\tilde 2}}{\tilde{m}_{2}^2-m_{3}^2}\\
0 & \frac{\sqrt{2}vA_{3\tilde 2}^{*}}{\tilde{m}_{2}^2-m_{3}^2} & 1-\frac{v^2|A_{3\tilde 2}|^2}{(m_{3}^2-\tilde{m}_{2}^2)^2 }
\end{pmatrix}\,,\\
W^{-4/3}&\approx\begin{pmatrix}
1 & \frac{\sqrt{2}v^2 Y_{3 \tilde 1}^{*}}{\tilde{m}_{1}^2-m_{3}^2}\\
\frac{-\sqrt{2}v^2 Y_{3 \tilde 1}}{\tilde{m}_{1}^2-m_{3}^2}& 1
\end{pmatrix}\,.\nonumber
\label{Wexpanded}
\end{align}
The physical LQ masses then read
\begin{align}
\begin{aligned}
\left( M_a^{ - 1/3} \right)^2 &\approx \left(\! m_1^2\! +\! v^2\left( \!{Y_1} \!- \!\frac{|{A_{\tilde 21}}{|^2}}{{\tilde m_2^2 - m_1^2}} \right),\;\tilde m_2^2\!+\! {v^2}\left(\! {{Y_{\tilde 2}} \!+\! \frac{{|{A_{\tilde 21}}{|^2}}}{{\tilde m_2^2 - m_1^2}} \!+\! \frac{{|{A_{3\tilde 2}}{|^2}}}{{\tilde m_2^2 - m_3^2}}} \right),\;\right. \\  & 
\quad\left. m_3^2  + {v^2}\left(\! {Y_3} \!-\! \frac{{|{A_{3\tilde 2}}{|^2}}}{{\tilde m_2^2 - m_3^2}} \right) \!\right)_a, \\
{\left( {M_a^{  2/3}} \right)^2} &\approx \left( \!m_2^2 \!+\! {v^2}{Y_2},\;\tilde m_2^2 \!+\! {v^2}\left( {{Y_{\tilde 2\tilde 2}} \!+\! {Y_{\tilde 2}}\! +\! \frac{{2|{A_{3\tilde 2}}{|^2}}}{{\tilde m_2^2 - m_3^2}}} \right),\right. \\
&\quad\left.\;m_3^2 \!+\! {v^2}\left( \!{{Y_3}+Y_{33} \!-\! \frac{{2|{A_{3\tilde 2}}{|^2}}}{{\tilde m_2^2 - m_3^2}}} \right)\! \right)_a ,\\
{\left( {M_a^{ - 4/3}} \right)^2} &\approx \left( {\tilde m_1^2 + {v^2}{Y_{\tilde 1}},\;m_3^2 + {v^2}({Y_3}-Y_{33})} \right)_a ,\\
{\left( {M^{5/3}} \right)^2} &\approx m_2^2 + {v^2}\left( {{Y_{22}} + {Y_2}} \right) \,,
\end{aligned}
\end{align}
valid up to order $v^2$, where $a$ runs from 1~to~3 for $Q=-1/3$ and $Q=2/3$ and from 1~to~2 for $Q=-4/3$, respectively.\footnote{For the calculation of the T parameter, we even needed the expansion of the mixing matrices and masses up to order $v^4$. However, these equations are too lengthy to be included in this work explicitly.}
\smallskip

We now write the interaction terms of the Higgs with the LQs in the form 
\begin{align}
\begin{aligned}
\mathcal{L}_{H\Phi}=&-\tilde{\Gamma}^{-1/3}_{ab}h\hat{\Phi}_{a}^{-1/3\,\dagger}\hat{\Phi}_{b}^{-1/3}-\tilde{\Gamma}^{2/3}_{ab}h\hat{\Phi}_{a}^{2/3\,\dagger}\hat{\Phi}_{b}^{2/3}-\tilde{\Gamma}^{-4/3}_{ab}h\hat{\Phi}_{a}^{-4/3\dagger}\hat{\Phi}_{b}^{-4/3}\\
&-\Gamma^{5/3}h\hat{\Phi}^{5/3\,\dagger}\hat{\Phi}^{5/3}-\tilde{\Lambda}^{-1/3}_{ab}h^2\hat{\Phi}_{a}^{-1/3\,\dagger}\hat{\Phi}_{b}^{-1/3}-\tilde{\Lambda}^{2/3}_{ab}h^2\hat{\Phi}_{a}^{2/3\,\dagger}\hat{\Phi}_{b}^{2/3}\\
&-\tilde{\Lambda}^{-4/3}_{ab}h^2\hat{\Phi}_{a}^{-4/3\,\dagger}\hat{\Phi}_{b}^{-4/3}-\Lambda^{5/3}h^2\hat{\Phi}^{5/3\,\dagger}\hat{\Phi}^{5/3}\,,
\end{aligned}
\end{align}
with $h$ as the physical Higgs field, $\hat \Phi^Q$ being the mass eigenstates of charge $Q$ with $a,b$ again running from 1~to~3 for $Q=-1/3$ and $Q=2/3$ and from 1~to~2 for $Q=-4/3$. In particular we have
\begin{align}
\begin{aligned}
\tilde{\Gamma}^{-1/3}&=W^{-1/3}\Gamma^{-1/3}W^{-1/3\,\dagger}\,,
&\quad\quad&\tilde{\Lambda}^{1/3}=W^{-1/3}\Lambda^{-1/3}W^{-1/3\,\dagger}\,,\\
\tilde{\Gamma}^{2/3}&=W^{2/3}\Gamma^{2/3}W^{2/3\,\dagger}\,,
&&\tilde{\Lambda}^{2/3}=W^{2/3}\Lambda^{2/3}W^{2/3\,\dagger}\,,\\
\tilde{\Gamma}^{-4/3}&=W^{-4/3}\Gamma^{-4/3}W^{-4/3\,\dagger}\,,
&&\tilde\Lambda^{-4/3}=W^{-4/3}\Lambda^{-4/3}W^{-4/3\,\dagger}\,,
\end{aligned}
\label{eq:higgs_couplings}
\end{align}
with
\begin{align}
\begin{aligned}
\Gamma^{-1/3}&=\frac{1}{\sqrt{2}}\begin{pmatrix}2vY_{1}&A_{\tilde{2}1}^{*}&2vY_{31}\\ A_{\tilde{2}1}&2vY_{\tilde{2}} & A_{3\tilde{2}} \\ 2vY_{31}^{*}& A_{3\tilde{2}}^{*}& 2vY_{3}\end{pmatrix}\,,
&&
\Lambda^{-1/3}=\frac{1}{2}\begin{pmatrix}Y_{1} & 0 & Y_{31}\\ 0& Y_{\tilde{2}} & 0\\ Y_{31}^{*} & 0 & Y_{3}\end{pmatrix}\,,\\
\Gamma^{2/3}&=\frac{1}{\sqrt{2}}\begin{pmatrix}2vY_{2} & 2vY_{\tilde{2}2}^{*} & 0\\ 2vY_{\tilde{2}2}& 2v\big(Y_{\tilde{2}}+Y_{\tilde{2}\tilde{2}}\big) &-\sqrt{2}A_{3\tilde{2}}^{*}\\ 0 &- \sqrt{2}A_{3\tilde{2}} & 2v\big(Y_{3}\!+\!Y_{33}\big)\end{pmatrix}\,, &&
\Lambda^{2/3}=\frac{1}{2}\begin{pmatrix}Y_{2} & Y_{\tilde{2}2}^{*} & 0\\ Y_{\tilde{2}2} & Y_{\tilde{2}}\!+\!Y_{\tilde{2}\tilde{2}} & 0\\ 0& 0& Y_{3}\!+\!Y_{33}\end{pmatrix}\,,\\
\Gamma^{-4/3}&=\frac{1}{\sqrt{2}}\begin{pmatrix}2vY_{\tilde{1}} & 2\sqrt{2}vY_{3\tilde{1}}^{*} \\ 2\sqrt{2}vY_{3\tilde{1}} & 2v(Y_{3}\!-\!Y_{33})\end{pmatrix}\,,&&
\Lambda^{-4/3}=\frac{1}{2}\begin{pmatrix}Y_{\tilde{1}} & \sqrt{2}Y_{3\tilde{1}}^{*} \\ \sqrt{2}Y_{3\tilde{1}} & Y_{3}\!-\!Y_{33}\end{pmatrix}\,.\\
\Gamma^{5/3}&=\sqrt{2}v\big(Y_{22}+Y_{2}\big)\,,
&&\Lambda^{5/3}=\frac{1}{2}\big(Y_{22}+Y_{2}\big)\,.
\end{aligned}
\end{align}
The expanded expressions for $\tilde{\Gamma}^{Q}$ and $\tilde{\Lambda}^{Q}$ up to $\mathcal{O}(v^2)$ are given in the appendix.
\smallskip

\section{Oblique Corrections}
\label{sec:Oblique}

Oblique Corrections, i.e. radiative corrections to the EW breaking sector of the SM, can be parametrized via the Peskin-Takeuchi parameters $S$, $T$ and $U$~\cite{Peskin:1990zt}. These parameters are expressed and calculated in terms of the vacuum polarization functions $\Pi_{VV}(q^2)$, with $V=W,Z,\gamma$. We use the convention
\begin{equation}
	\begin{gathered}
		\begin{overpic}[scale=.45,,tics=10]
		{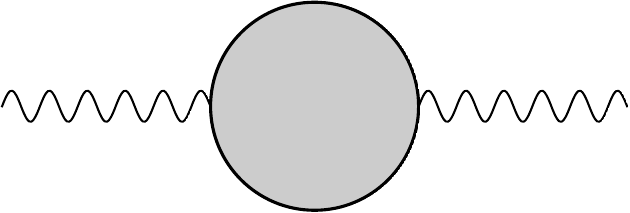}
		\put(7,23){$V^\mu$}
		\put(82,23){$V^\nu$}
		\end{overpic}
	\end{gathered}
	\hspace{0.3cm}
	=i\Pi_{VV}(q^2)g^{\mu\nu}-i\Delta(q^2)q^{\mu}q^{\nu} \ .
\end{equation}
Taking into account that our NP scale is higher than the EW breaking scale, we can expand the gauge bosons self-energies in $q^2/M^2$. As $\Delta(q^2)$ has no physical effect, the three oblique parameters can be written as
\begin{align}
\begin{split}
 S&=-\frac{4s_{w}^2 c_{w}^2}{\alpha m_{Z}^2}\bigg(\Pi_{ZZ}(0)-\Pi_{ZZ}(m_{Z}^2)+\Pi_{\gamma\gamma}(m_{Z}^2)+\frac{c_w^2-s_w^2}{c_{w}s_{w}}\Pi_{Z\gamma}(m_{Z}^2)\bigg)\,,\\
 T&=\frac{\Pi_{WW}(0)}{\alpha m_{W}^2}-\frac{\Pi_{ZZ}(0)}{\alpha m_{Z}^2}\,,\\
 U&=-\frac{4s_{w}^2c_{w}^2}{\alpha}\bigg(\frac{\Pi_{WW}(0)-\Pi_{WW}(m_{W}^2)}{c_{w}^2m_{W}^2}-\frac{\Pi_{ZZ}(0)-\Pi_{ZZ}(m_{Z}^2)}{m_{Z}^2}\\
&\phantom{=4s_{w}^2c_{w}^{2}123}+\frac{s_{w}^2}{c_{w}^2}\frac{\Pi_{\gamma\gamma}(m_{Z}^2)}{m_{Z}^2}+2\frac{s_{w}}{c_{w}}\frac{\Pi_{Z\gamma}(m_{Z}^2)}{m_{Z}^2}\bigg)\,,
\end{split}
\end{align}
where we used renormalization conditions for the vector fields such that
\begin{align}
\begin{aligned}
\Pi_{\gamma\gamma}(0)=\Pi_{Z\gamma}(0)=\text{Re}\big[\Pi_{ZZ}(m_{Z}^2)\big]=\text{Re}\big[\Pi_{WW}(m_{W}^2)\big]=0\,.
\end{aligned}
\end{align}
These conditions are fulfilled automatically for $\Pi_{\gamma\gamma}$ and $\Pi_{Z\gamma}$ because of the Ward identities.

\begin{figure}
	\centering
	\begin{overpic}[scale=.43,,tics=10]
		{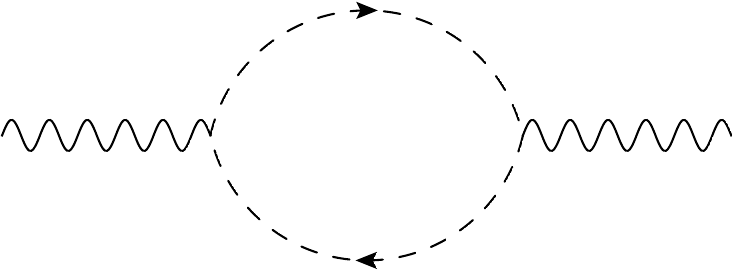}
		\put(5,24){$V^{\mu}$}
		\put(88,24){$V^{\nu}$}
		\put(46,40){$\hat{\Phi}_{a}^{Q^{\prime}}$}
		\put(46,6){$\hat{\Phi}_{b}^Q$}
	\end{overpic}
\hfill
	\begin{overpic}[scale=.5,,tics=10]
		{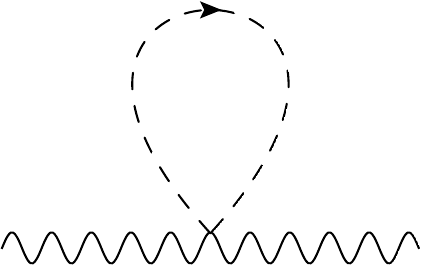}
		\put(5,13){$V^{\mu}$}
		\put(88,13){$V^{\nu}$}
		\put(45,66){$\hat{\Phi}_{a}^{Q}$}
	\end{overpic}
\hfill
	\begin{overpic}[scale=.46,,tics=10]
		{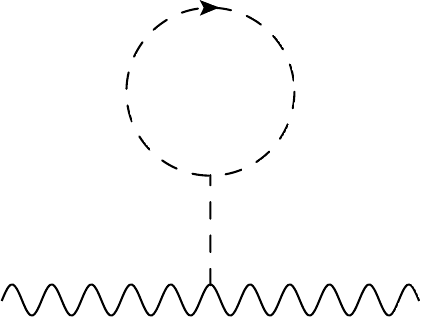}
		\put(5,13){$V^{\mu}$}
		\put(88,13){$V^{\nu}$}
		\put(45,79){$\hat{\Phi}_{a}^{Q}$}
		\put(54,16){$h$}
	\end{overpic}
	\caption{The three different topologies of Feynman diagrams that contribute to $\Pi_{VV}(q^2)$ with $V=W,Z,\gamma$. The last diagram only exists for $V=W,Z$ and has no impact on the $S$, $T$ and $U$ parameters as it is momentum independent.}
	\label{fig:diagramm_VV_SE}
\end{figure}
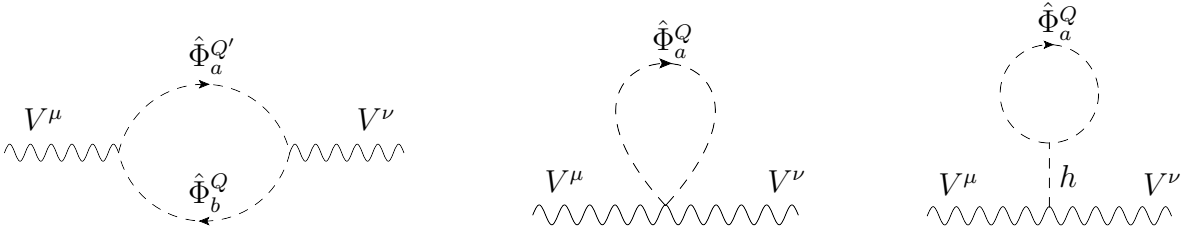

$S$, $T$ and $U$ can be calculated with the bare (unrenormalized) two-point correlation functions, the corresponding diagrams in our model are shown in Fig.~\ref{fig:diagramm_VV_SE}. Therefore, we used the check that all divergences disappear in the physical observables $S$, $T$ and $U$ after having summed over all $SU(2)_{L}$ components in the loop. The complete expressions for these parameters are quite lengthy and therefore given in the appendix. Expanding in addition in $q^2/M^2$ and in $v/M$, i.e. perturbatively diagonalizing the LQ mass matrices, we can however obtain relatively compact expressions. Up to leading order in $v$ we find
\begin{align}
S&\approx-\frac{N_{c}\, v^2}{36 \pi} \bigg(\frac{7Y_{22}}{m_{2}^2}+\frac{Y_{\tilde{2}\tilde{2}}}{\tilde{m}_{2}^2}-\frac{8Y_{33}}{m_{3}^2}-\frac{|A_{\tilde 21}|^2}{10\tilde{m}_{2}^4}\mathcal{K}_{1}\Big(\frac{m_{1}^2}{\tilde{m}_{2}^2}\Big)+\frac{17|A_{3\tilde 2}|^2}{10\tilde{m}_{2}^4}\mathcal{K}_{2}\Big(\frac{m_{3}^2}{\tilde{m}_{2}^2}\Big)\bigg)\,,\nonumber\\
T&\approx \frac{N_{c}v^2}{24\pi g_{2}^2 s_{w}^2}\bigg(\frac{Y_{22}^2}{m_{2}^2}+\frac{Y_{\tilde{2}\tilde{2}}^2}{\tilde{m}_{2}^2}+\frac{4Y_{33}^2}{m_{3}^2}+\frac{|A_{\tilde{2}1}|^4}{10\tilde{m}_{2}^6}\mathcal{K}_{3}\Big(\frac{m_{1}^2}{\tilde{m}_{2}^2}\Big)+\frac{|A_{3\tilde{2}}|^4}{2\tilde{m}_{2}^6}\mathcal{K}_{4}\Big(\frac{m_{3}^2}{\tilde{m}_{2}^2}\Big)+\frac{Y_{\tilde{2}\tilde{2}}|A_{\tilde{2}1}|^2}{2\tilde{m}_{2}^4}\mathcal{K}_{5}\Big(\frac{m_{1}^2}{\tilde{m}_{2}^2}\Big)\nonumber\\
&-\frac{Y_{\tilde{2}\tilde{2}}|A_{3\tilde{2}}|^2}{2\tilde{m}_{2}^4}\mathcal{K}_{5}\Big(\frac{m_{3}^2}{\tilde{m}_{2}^2}\Big)-2\frac{Y_{33}|A_{3\tilde{2}}|^2}{\tilde{m}_{2}^4}\mathcal{K}_{6}\Big(\frac{m_{3}^2}{\tilde{m}_{2}^2}\Big)+\frac{4|Y_{31}|^2}{m_{3}^2}\mathcal{K}_{7}\Big(\frac{m_{1}^2}{m_{3}^2}\Big)-\frac{4|Y_{3\tilde{1}}|^2}{m_{3}^2}\mathcal{K}_{7}\Big(\frac{\tilde{m}_{1}^2}{m_{3}^2}\Big)\nonumber\\&-\frac{2|Y_{2\tilde{2}}|^2}{\tilde{m}_{2}^2}\mathcal{K}_{7}\Big(\frac{m_{2}^2}{\tilde{m}_{2}^2}\Big)-\frac{2\Re\big[Y_{31}A_{\tilde{2}1}A_{3\tilde{2}}^{*}\big]}{\tilde{m}_{2}^4}\mathcal{K}_{8}\Big(\frac{m_{1}^2}{\tilde{m}_{2}^2},\frac{m_{3}^2}{\tilde{m}_{2}^2}\Big)+\frac{|A_{\tilde{2}1}|^2|A_{3\tilde{2}}|^{2}}{5\tilde{m}_{2}^6}\mathcal{K}_{9}\Big(\frac{m_{1}^2}{\tilde{m}_{2}^2},\frac{m_{3}^2}{\tilde{m}_{2}^2}\Big)\bigg)\,,\\
U&\approx 0\,,\nonumber
\label{eq:oblique_parameters}
\end{align}
where the loop functions, given in the appendix, are normalized to be unity in case of equal masses. These expressions agree with Refs.~\cite{Froggatt:1991qw,Keith:1997fv} for the special cases studied there. Note that $U$ is approximately zero since it only arises at dimension 8.
\bigskip

\section{Higgs Couplings to $g$, $\gamma$ and $Z$}
\label{sec:Higgs}

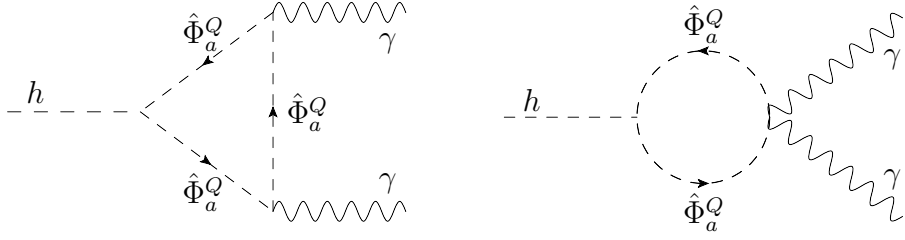
\begin{figure}
	\centering
	\begin{overpic}[scale=.5,,tics=10]
		{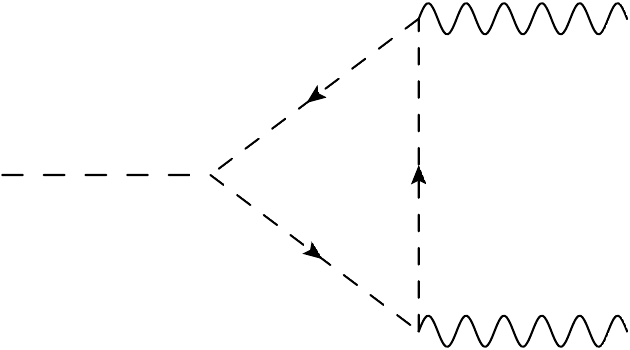}
		\put(5,30){$h$}
		\put(70,25){$\hat{\Phi}_{a}^{Q}$}
		\put(44,45){$\hat{\Phi}_{a}^{Q}$}
		\put(44,5){$\hat{\Phi}_{a}^Q$}
		\put(93,44){$\gamma$}
		\put(93,9){$\gamma$}
	\end{overpic}
	\hspace{1cm}
	\begin{overpic}[scale=.5,,tics=10]
		{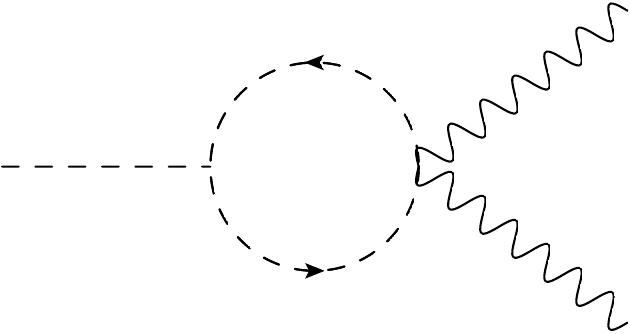}
		\put(5,28){$h$}
		\put(45,47){$\hat{\Phi}_{a}^{Q}$}
		\put(45,0){$\hat{\Phi}_{a}^{Q}$}
		\put(95,40){$\gamma$}
		\put(95,10){$\gamma$}
	\end{overpic}
	\caption{The two types of diagrams that induce NP effects in $h\to\gamma\gamma$. For $h\to gg$ the photons can simply be replaced by gluons, for $h \to Z \gamma$ one photon can be replaced by a $Z$~boson. The additional diagrams with reversed charge flow are not depicted.}
	\label{fig:diagramm_h_gammagamma}
\end{figure}

The Feynman diagrams involving scalar LQs contributing to $h\to\gamma\gamma$, $h\to gg$ and $h \to Z \gamma$ are shown in Fig.~\ref{fig:diagramm_h_gammagamma}. The amplitude, induced by them, reads
\begin{align}
\mathcal{A}\big[h\to\gamma(p_1)\gamma(p_2)\big]=\frac{\alpha N_{c}}{24\pi}\sum_{Q,a}\frac{Q^2\tilde{\Gamma}_{aa}^{Q}}{{(M_{a}^Q)}^2}\big(m_{h}^2\varepsilon(p_1)\!\cdot\!\varepsilon(p_2)-2(\varepsilon(p_1)\!\cdot\! p_{2})(\varepsilon(p_2)\!\cdot\! p_{1})\big)\,,
\end{align}
with $p_{1}$ and $p_{2}$ representing the photon momenta, $\varepsilon_{\mu}(p_i)$ the corresponding polarization vectors and $a$ running over the number of mass eigenstates with the same electric charge $Q=\{-1/3,\,2/3,\,-4/3,\,5/3\}$. Here we used on-shell kinematics and expanded in $m_h^2/M^2$. 
\smallskip

Similarly, for the decay into a pair of gluons, we obtain
\begin{align}
\mathcal{A}\big[h\to g^{A}(p_1)g^{A}(p_2)\big]=\frac{\alpha_s}{48\pi}\sum_{Q}\frac{\tilde{\Gamma}_{aa}^{Q}}{{(M_{a}^Q)}^2}\big(m_{h}^2\varepsilon^{A}(p_1)\!\cdot\!\varepsilon^{A}(p_2)-2(\varepsilon^{A}(p_1)\!\cdot\! p_{2})(\varepsilon^{A}(p_2)\!\cdot\! p_{1})\big)\,,
\nonumber
\end{align}
where $A$ labels the 8 gluons (no sum implied). For the Higgs decaying into a $Z$ and a photon we obtain
\begin{align}
\begin{split}
\mathcal{A}[h \to Z(p_Z) \gamma(p_\gamma)] = \frac{\alpha N_c}{24 \pi} \frac{1}{s_w c_w} \sum_{Q} \Bigg( Q\frac{\tilde{T}^Q_{ab}\,\tilde{\Gamma}^Q_{ba}}{{(M_b^Q)}^2} \,  \, \mathcal{K}_7\!\left(x_{ab}^Q\right) -s_w^2 Q^2 \frac{\tilde{\Gamma}^Q_{aa}}{{(M_a^Q)}^2}  \Bigg) \\ \times
\Big( (m_h^2-m_Z^2)\, \varepsilon(p_Z)\cdot \varepsilon(p_\gamma) -2 (\varepsilon(p_Z)\!\cdot\! p_\gamma) (\varepsilon(p_\gamma)\!\cdot\! p_Z) \Big) \,,
\end{split}
\end{align}
with a simultaneous expansion in $m_h^2/M^2$ and $m_Z^2/M^2$ and
\begin{align}
x_{ab}^Q=\frac{{(M_a^Q)}^2}{{(M_b^Q)}^2} \ .
\end{align}
\smallskip

The relevant observables in this context are the effective on-shell $h\gamma\gamma$, $hgg$ and $h Z\gamma$ couplings, normalized to their SM values
\begin{align}
\kappa_{\gamma}=\sqrt{\frac{\Gamma_{h\to\gamma\gamma}}{\Gamma_{h\to\gamma\gamma}^{\text{SM}}}}\,,&&
\kappa_{g}=\sqrt{\frac{\Gamma_{h\to gg}}{\Gamma_{h\to gg}^{\text{SM}}}}\,,&&\kappa_{Z\gamma}=\sqrt{\frac{\Gamma_{h\to Z\gamma}}{\Gamma_{h\to Z\gamma}^{\text{SM}}}}\,.
\end{align}
We then have
\begin{align}
\begin{aligned}
\kappa_{\gamma}&=1+\frac{1}{\mathcal{A}_{h\to\gamma\gamma}^{\text{SM}}}\frac{\alpha N_{c}}{24\pi} \sum_{Q}Q^2\frac{\tilde{\Gamma}_{aa}^{Q}}{{(M_{a}^Q)}^2}\,,\\
\kappa_{g}&=1+\frac{1}{\mathcal{A}_{h\to gg}^{\text{SM}}} \frac{\alpha_s}{48\pi}\sum_{Q}\frac{\tilde{\Gamma}_{aa}^{Q}}{{(M_{a}^Q)}^2}\,,\\
\kappa_{Z\gamma}&=1-\frac{1}{\mathcal{A}_{h\to Z\gamma}^{\text{SM}}}\frac{\alpha N_c}{24 \pi} \frac{1}{s_w c_w} \sum_{Q} \bigg( Q\frac{\tilde{T}^Q_{ab}\,\tilde{\Gamma}^Q_{ba}}{{(M_b^Q)}^2} \,  \, \mathcal{K}_6\!\left(x_{ab}^Q\right) -s_w^2 Q^2\frac{\tilde{\Gamma}^Q_{aa}}{{(M_a^Q)}^2} \bigg)\,,
\end{aligned}
\end{align}
with the LO SM amplitudes (see e.g. Ref.~\cite{Spira:2016ztx} for an overview) given by~\cite{Ellis:1975ap,Cahn:1978nz,Bergstrom:1985hp,Inami:1982xt,Djouadi:1991tka,Spira:1995rr,Dorsner:2016wpm}
\begin{align}
\begin{aligned}
\mathcal{A}_{h\to\gamma\gamma}^{\text{SM}}&=\frac{\alpha}{4\pi \sqrt{2}v}\Big(A_{1}(x_W)+\frac{4}{3}A_{1/2}(x_t)\Big)\,,\\
\mathcal{A}_{h\to gg}^{\text{SM}}&=\frac{\alpha_s}{8\pi \sqrt{2}v}A_{1/2}(x_{t})\,,\\
\mathcal{A}_{h\to Z\gamma}^{\text{SM}}&=\frac{\alpha}{4\pi s_{w}\sqrt{2}v}\Big(c_{w}C_{1}(x_{W}^{-1},y_{W})+\frac{2}{c_{w}}\Big(1-\frac{8}{3}s_{w}^2\Big)C_{1/2}(x_{t}^{-1},y_{t})\Big)\,.
\label{eq:SM_amplitude_Higgs}
\end{aligned}
\end{align}
We defined
\begin{align}
x_{i}=\frac{m_{h}^2}{4m_{i}^2}\,,&&y_{i}=\frac{4m_{i}^2}{m_{Z}^2}\,,
\end{align}
while the loop functions are given in the appendix.\footnote{Note that we did not include the effects of bottom quarks in the SM prediction which would lead to a 10\% destructive interference.}
\bigskip

\begin{figure}
	\centering
	\includegraphics[width=0.8\textwidth]{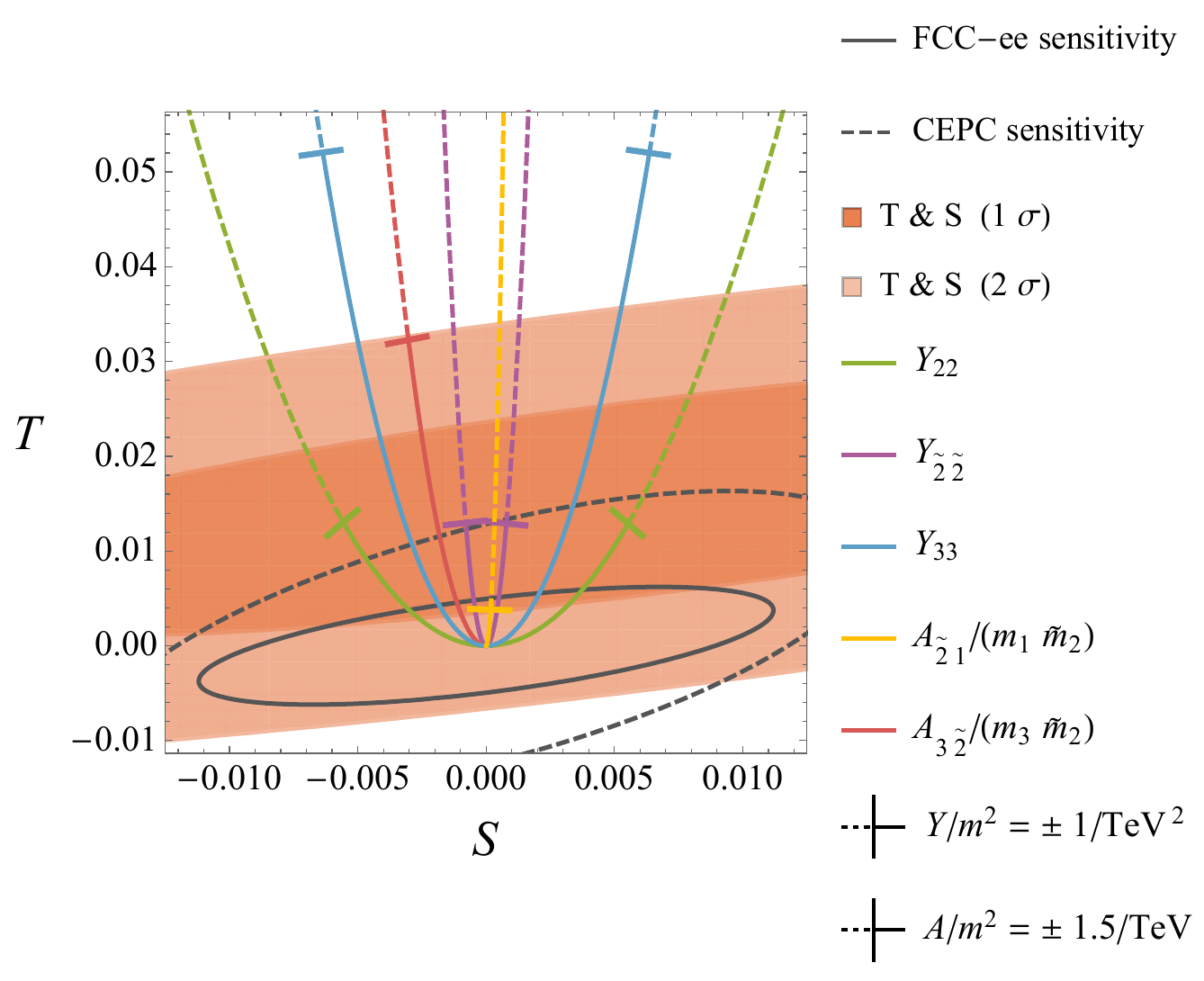}
	\caption{Correlations between $S$ and $T$ for four different Lagrangian parameters in Eq. \eqref{eq:Higgs_LQ_lagrangian}, assuming that only one of them is non-zero at a time. For simplicity, we assumed all LQ masses to be equal. While $Y_{22}$ and $Y_{\tilde{2}\tilde{2}}$ can yield both positive and negative effects in $S$, the effect in the $T$ parameter is positive definite. Since our prediction for $S$ and $T$ depends on a single combination of parameters ($Y/m^2$ or $A^2/m^4$), we used one degree of freedom to obtain the preferred region in the $S$-$T$ plane, such that the region within the ellipse labelled by $1\,\sigma$ ($2\,\sigma$) corresponds to $68\%$ C.L. ($95\%$ C.L.).}
	\label{STcorrelations}	
\end{figure}

\begin{figure}
	\centering
	\includegraphics[width=0.8\textwidth]{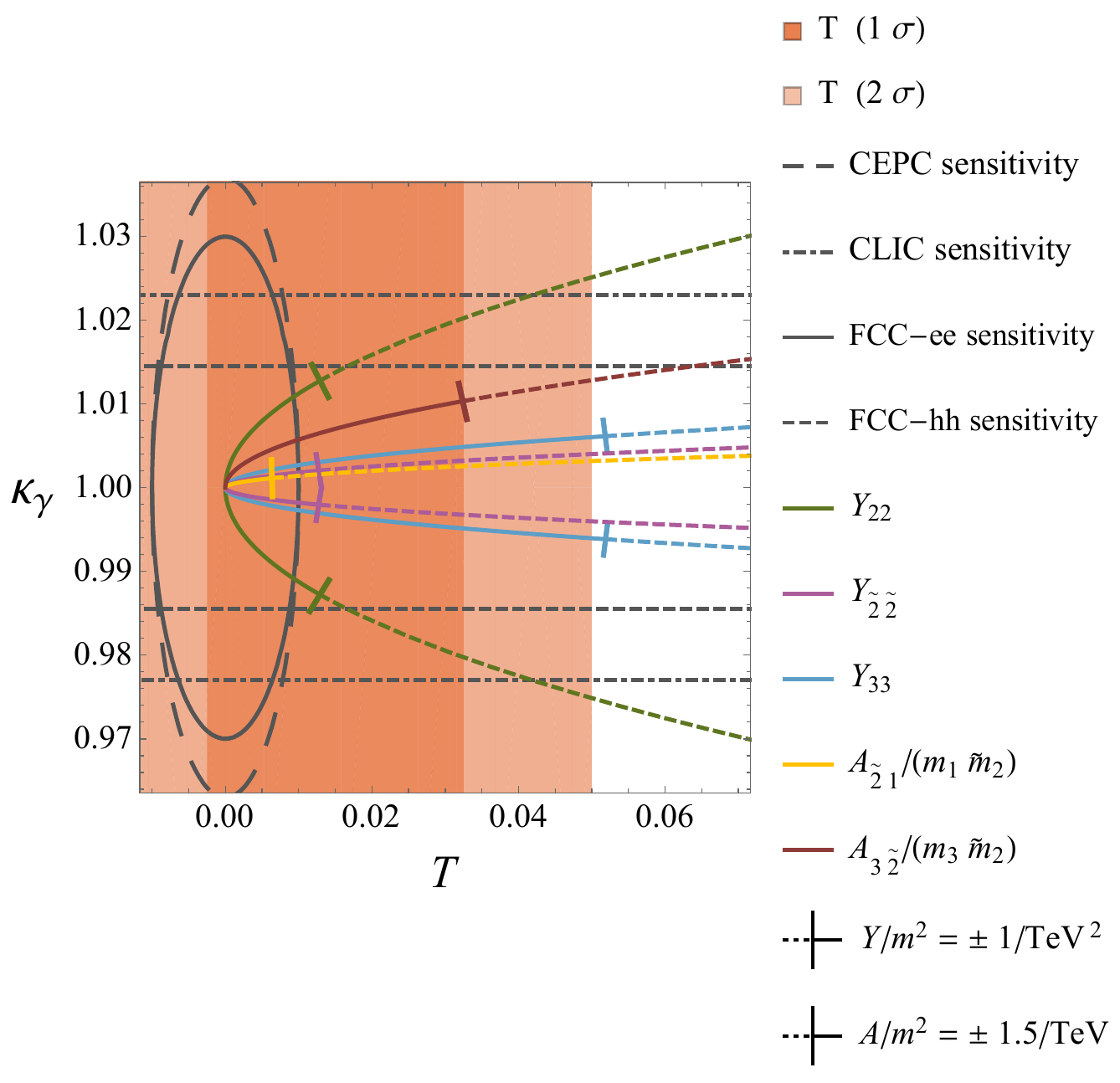}
	\caption{Correlations between $\kappa_\gamma$ and $T$ for different Lagrangian parameters, assuming that only one of them is non-zero at a time and assuming all LQ masses to be equal.}
	\label{kappaTkorrelations}	
\end{figure}

In addition to the expansion of the loop functions, we can also expand the expressions $\tilde{\Gamma}^{Q}/M^2$ and $\tilde{T}^Q \,\tilde{\Gamma}^{Q} \, \mathcal{K}_6(x_{ab}^Q) /M^2$ in $v^2/M^2$ up to $\mathcal{O}(v^3)$, using \eq{Wexpanded}. We obtain
\begin{align}
\sum_{a=1}^3\frac{\tilde{\Gamma}^{-1/3}_{aa}}{{(M_{a}^{-1/3})}^2}&\approx \sqrt{2}v\bigg(\frac{Y_{1}}{m_{1}^2}+\frac{Y_{\tilde{2}}}{\tilde{m}_{2}^2}+\frac{Y_{3}}{m_{3}^2}-\frac{|A_{\tilde 21}|^2}{m_{1}^2\tilde{m}_{2}^2}-\frac{|A_{3\tilde 2}|^{2}}{m_{3}^2\tilde{m}_{2}^2}\bigg)\,,
\nonumber\\
\sum_{a=1}^3\frac{\tilde{\Gamma}^{2/3}_{aa}}{{(M_{a}^{2/3})}^2}&\approx\sqrt{2}v\bigg(\frac{Y_{2}}{m_{2}^2}+\frac{Y_{\tilde{2}\tilde{2}}+Y_{\tilde{2}}}{\tilde{m}_{2}^2}+\frac{Y_{3}+Y_{33}}{m_{3}^2}-\frac{2|A_{3\tilde 2}|^{2}}{\tilde{m}_{2}^2 m_{3}^2}\bigg)\,,
\nonumber\\
\sum_{a=1}^2\frac{\tilde{\Gamma}^{-4/3}_{aa}}{{(M_{a}^{-4/3})}^2}&\approx\sqrt{2}v\bigg(\frac{Y_{\tilde{1}}}{\tilde{m}_{1}^2}+\frac{Y_{3}-Y_{33}}{m_{3}^2}\bigg)\,,
\nonumber\\
\frac{\Gamma^{5/3}}{{(M^{5/3})}^2}&\approx\sqrt{2}v\frac{Y_{22}+Y_{2}}{m_{2}^2}\,,
\end{align}

\begin{align}
\sum_{a,b=1}^3\frac{\tilde{T}^{-1/3}_{ab}\,\tilde{\Gamma}^{-1/3}_{ba}}{{(M_b^{-1/3})}^2} \,  \, \mathcal{K}_7\big(x_{ab}^{-1/3}\big) &\approx \frac{v}{\sqrt{2}}\! \left(\frac{|A_{21}|^2}{2\tilde{m}_2^4} \mathcal{F}_1\!\left(\frac{m_1^2}{\tilde{m}_2^2}\right) + \frac{|A_{32}|^2}{2\tilde{m}_2^4}\mathcal{F}_1\!\left(\frac{m_3^2}{\tilde{m}_2^2}\right) -\frac{Y_{\tilde{2}}}{\tilde{m}_2^2} \right) \, , 
\nonumber\\
\sum_{a,b=1}^3\frac{\tilde{T}^{2/3}_{ab}\,\tilde{\Gamma}^{2/3}_{ba}}{{(M_b^{2/3})}^2} \,  \, \mathcal{K}_7\big(x_{ab}^{2/3}\big) &\approx \frac{v}{\sqrt{2}}\! \left(  \frac{Y_{\tilde{2}}}{\tilde{m}_2^2} - \frac{Y_2}{m_2^2}  + \frac{2(Y_3+Y_{33})}{m_3^2} + \frac{Y_{\tilde{2}\tilde{2}}}{\tilde{m}_2^2} - \frac{3|A_{32}|^2}{\tilde{m}_2^4} \mathcal{F}_2\!\left(\frac{m_3^2}{\tilde{m}_2^2}\right)\!\right) \, ,
\nonumber\\
\sum_{a,b=1}^2\frac{\tilde{T}^{-4/3}_{ab}\,\tilde{\Gamma}^{-4/3}_{ba}}{{(M_b^{-4/3})}^2} \,  \, \mathcal{K}_7\big(x_{ab}^{-4/3}\big) &\approx - \sqrt{2} v \frac{Y_3-Y_{33}}{m_3^2} \, ,
\nonumber \\
\frac{\tilde{T}^{5/3}\,\tilde{\Gamma}^{5/3}}{{(M^{5/3})}^2}  &\approx \frac{v}{\sqrt{2}} \frac{Y_{22} + Y_{2}}{m_2^2}  \ .
\end{align}
Therefore, we have directly expressed $\kappa_{\gamma}$, $\kappa_{g}$ and $\kappa_{Z\gamma}$ in terms of the Lagrangian parameters. The loop functions $\mathcal{F}_1$ and $\mathcal{F}_2$, given in the appendix, are again normalized to be unity in case of equal masses. 
\smallskip

\begin{figure}
	\renewcommand\tabularxcolumn[1]{m{#1}}
	\setkeys{Gin}{width=1\linewidth, 
		height=1\linewidth,
		keepaspectratio}
	\centering
	\resizebox{\textwidth}{!}{
	\begin{tabularx}{1.12\linewidth}{@{} XX @{}}
		\qquad
		\subfloat{\includegraphics[width=\linewidth]{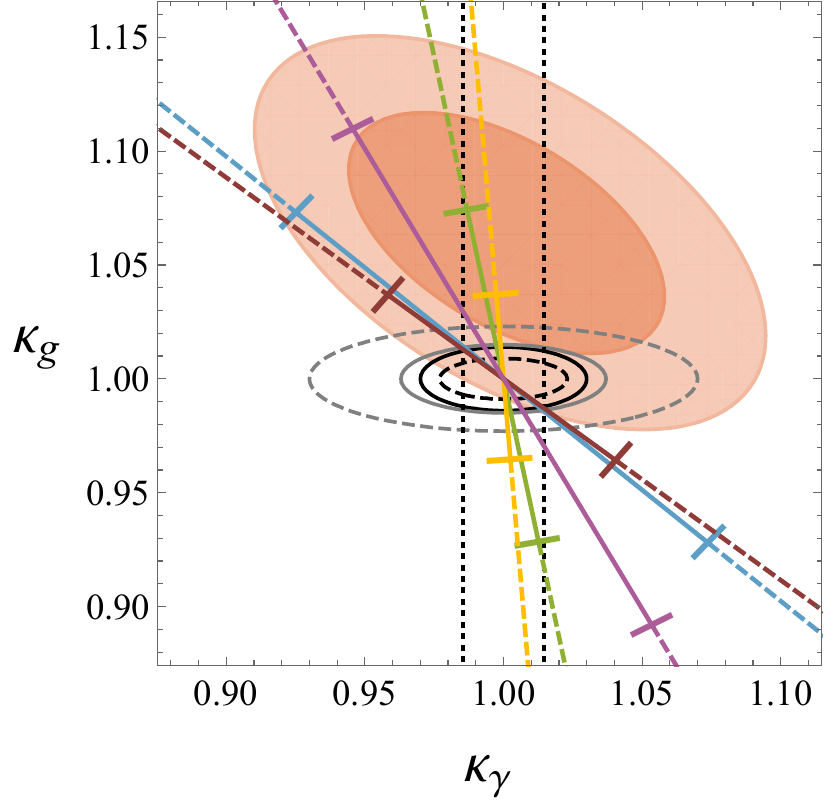}}
		
		\vspace{0.2cm}
		\qquad
		\subfloat{\includegraphics[width=\linewidth]{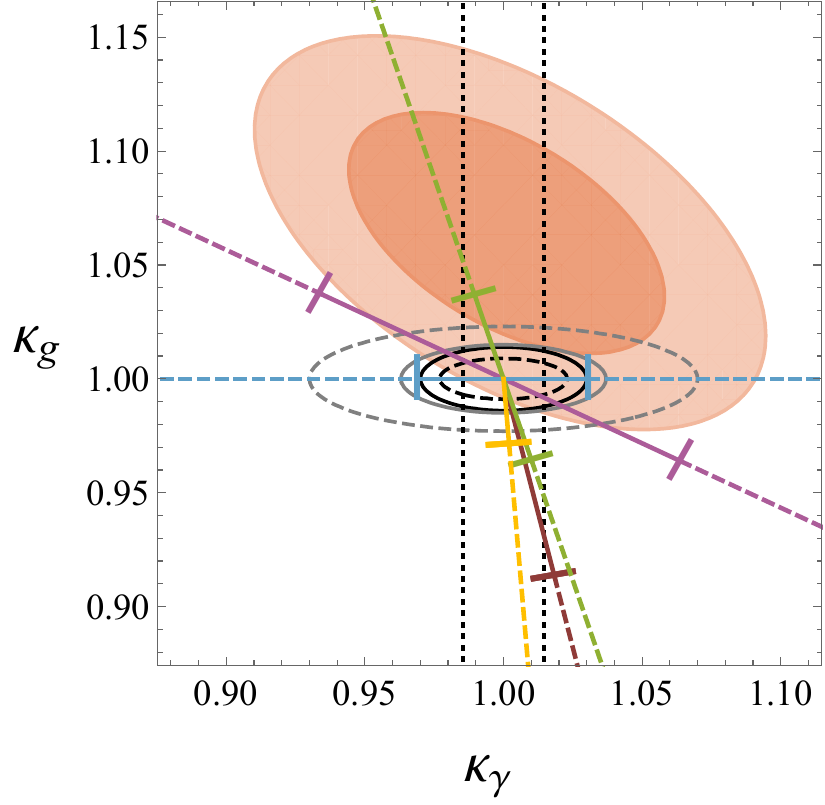}}
		
		&
		
		\qquad\quad
		\subfloat{\includegraphics[width=1.4\linewidth]{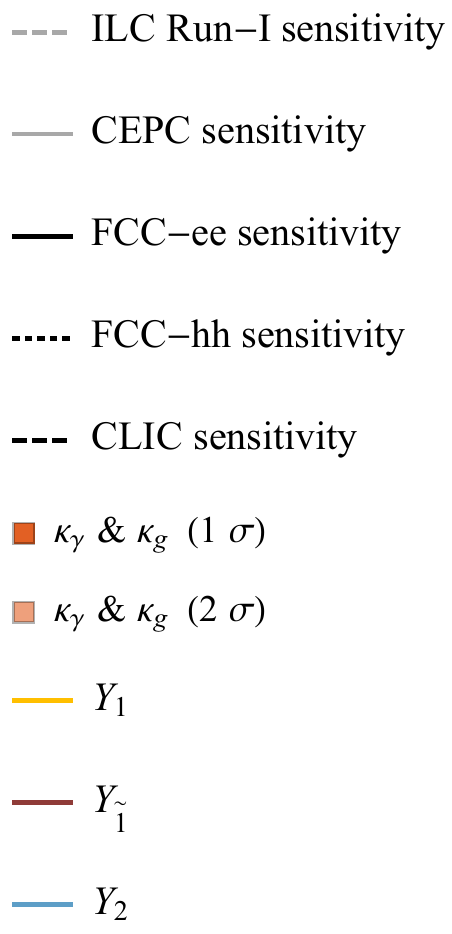}}
		
		\qquad\quad
		\subfloat{\includegraphics[width=1.4\linewidth]{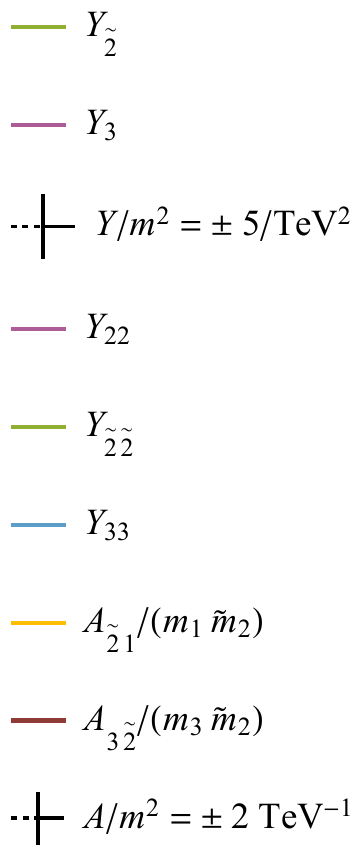}}
		
	\end{tabularx}
	}
	\caption{Correlations between $\kappa_\gamma$ and $\kappa_g$ for the different Lagrangian parameters. Here we assumed all bi-linear LQ mass terms to be equal. Here we used one degree of freedom in the $\chi^2$ fit for the allowed regions and the future prospects such that the intersection with the LQ line indicates the $68\%$ and $95\%$ CL for the corresponding parameter $Y/m^2$ or $A^2/m^4$.}
	\label{kappacorrelations}
\end{figure}

\section{Phenomenological Analysis}
\label{sec:Pheno}

Before we illustrate the effects of LQs in the observables of our interest, let us recall the current experimental situation and the prospects at future colliders. Concerning the oblique corrections, the global fit to electroweak precision measurements (including LEP~\cite{ALEPH:2005ab}, Tevatron~\cite{Aaltonen:2013iut} and LHC~\cite{Aad:2019bdc}) of Ref.~\cite{Ellis:2018gqa} constrains the $S$ and $T$ parameter to lie within
\begin{align}
S=[-0.06,0.07]~, && T=[-0.02,0.05]\,,
\end{align}
at $95\%$ C.L. within the 2-dimensional $S$-$T$ plane, with a correlation factor of $0.72$. Here, we can optimistically expect a sensitivity of $0.008$ in the future at the FCC-ee~\cite{Abada:2019zxq}.

For on-shell Higgs couplings, we used the results of Refs.~\cite{Bernon:2015hsa,Kraml:2019sis} for the current status, which are 
\begin{align}
\kappa_{g}=1.066\substack{+0.051\\-0.050}~,&&\kappa_{\gamma}=0.999\substack{+0.055\\-0.053}\,.
\end{align}
Concerning future prospects we expect for $\kappa_{\gamma}$ ($\kappa_g$) an accuracy of 7\% (2.3\%) at the ILC~\cite{Behnke:2013lya}, 3.7\% (1.5\%) at CEPC~\cite{An:2018dwb}, 2.3\% (0.9\%) at CLIC~\cite{Aicheler:2012bya}, 3\% (1.4\%) at the FCC-ee~\cite{Abada:2019zxq} and 1.45\% at the FCC-hh~\cite{Benedikt:2018csr}. Finally, concerning $h\to Z\gamma$, an accuracy of up to $1.8\%$ in $h\to Z\mu^+\mu^-/h\to \mu^+\mu^-$ can be achieved at the FCC-ee~\cite{Abada:2019zxq}.

Let us start by considering the oblique parameters. Here and in the following, we will for definiteness assume a LQ mass of $1\,$TeV, which is compatible with current LHC limits~\cite{Sirunyan:2018ruf,Aaboud:2019jcc,Aaboud:2019bye} for a broad range of couplings to fermions. In Fig.~\ref{STcorrelations} we show the correlations between $S$ and $T$ for the four cases which contribute to both parameters simultaneously. As one can see, the effect in $T$ is positive definite, as slightly preferred by current data. Note that the $A$ parameters are dimensionful couplings which are naturally expected to be of the same order as the LQ masses and that similarly the dimensionless couplings $Y$ are expected to be of order 1. Therefore, $T$ already now sets relevant limits on these couplings and its future experimental sensitivity allows for stringent constraints or even to discover deviations from the SM within LQ models.

\begin{figure}
	\renewcommand\tabularxcolumn[1]{m{#1}}
	\setkeys{Gin}{width=1\linewidth, 
		height=1\linewidth,
		keepaspectratio}
	\centering
	\resizebox{\textwidth}{!}{
	\begin{tabularx}{1.15\linewidth}{@{} XX @{}}
		\qquad
		\subfloat{\includegraphics[width=1\linewidth]{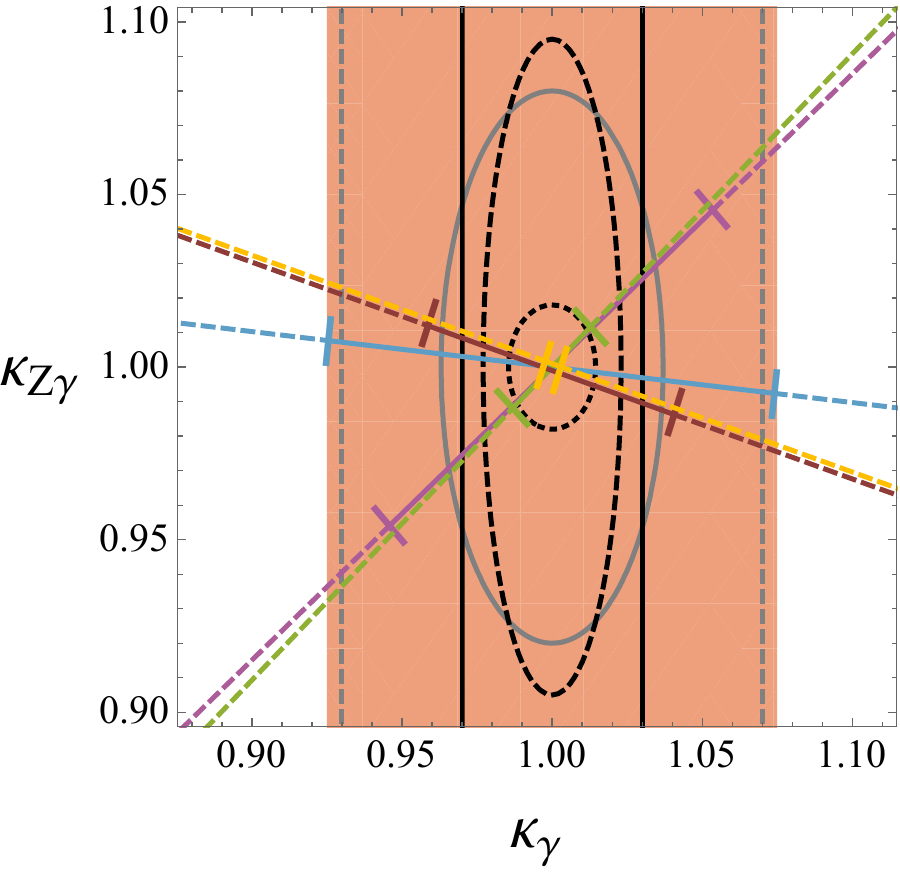}}
		
		\vspace{0.2cm}
		\qquad
		\subfloat{\includegraphics[width=1\linewidth]{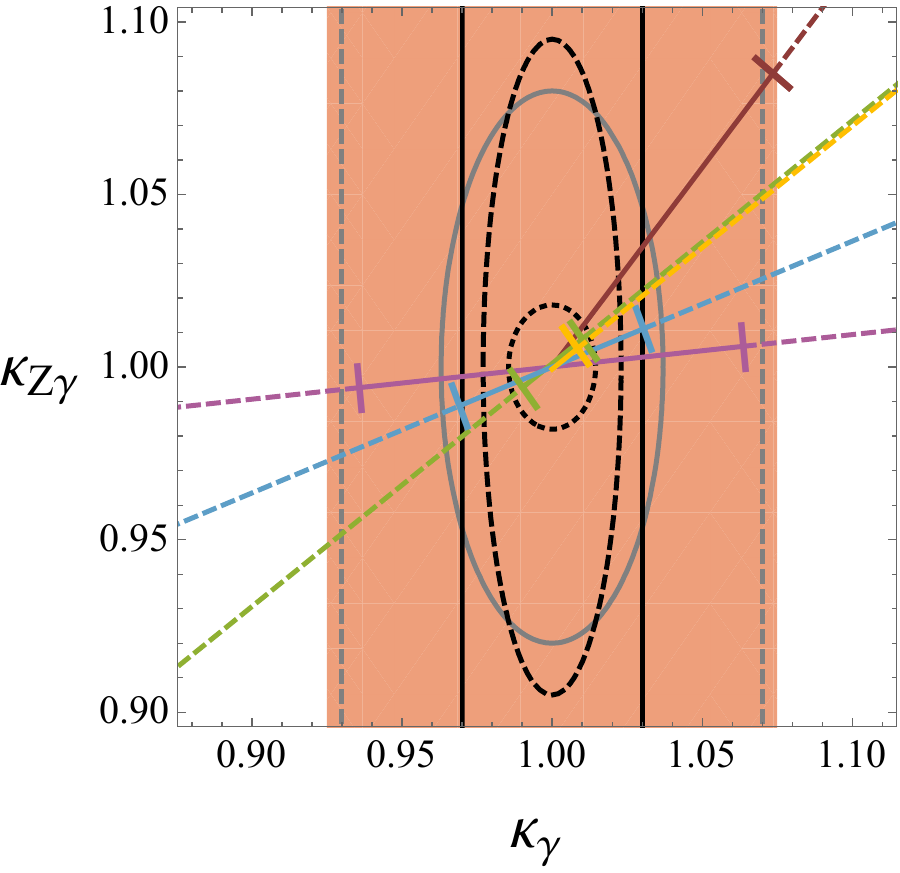}}
		
		&
		
		\qquad\quad
		\subfloat{\includegraphics[width=1\linewidth]{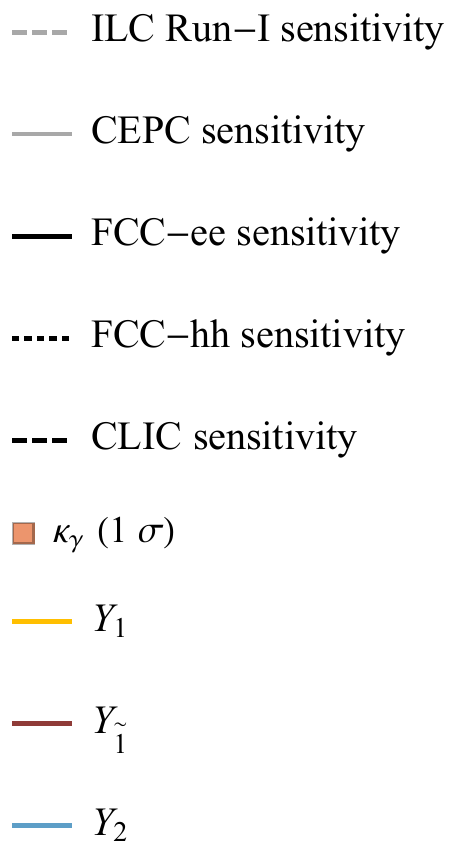}}
		
		\qquad\quad
		\subfloat{\includegraphics[width=1\linewidth]{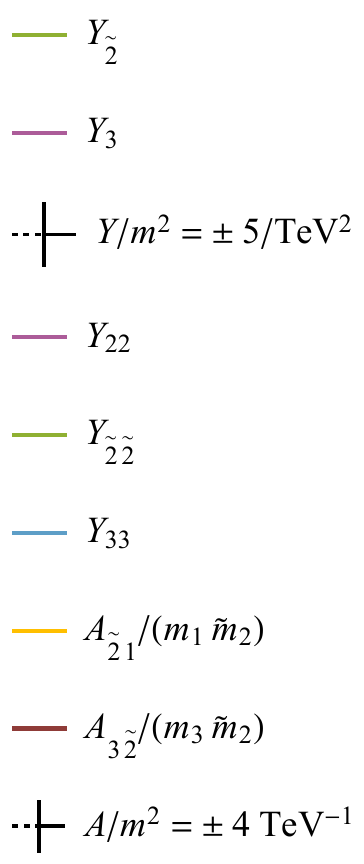}}
		
	\end{tabularx}
	}
	\caption{Correlations between $\kappa_\gamma$ and $\kappa_{Z\gamma}$ for the different Lagrangian parameters coupling LQs to the Higgs. The currently preferred regions are shown as red ellipses and the future sensitivity is indicated by the dashed and dotted lines.}
	\label{Zgamma}
\end{figure}

Turning to the effects in Higgs couplings to gauge bosons, we show the correlations between $\kappa_\gamma$ and $T$ in Fig.~\ref{kappaTkorrelations} and between $\kappa_\gamma$ and $\kappa_g$ in Fig.~\ref{kappacorrelations}. The currently allowed regions ($1\,\sigma$ and $2\,\sigma$, corresponding to $68\%$ and $95\%$ C.L. for one degree of freedom) are shown in color while the future prospects are indicated by dashed and dotted boundaries of the corresponding ellipses. Assuming a value close to the current best fit point in the $\kappa_\gamma$-$\kappa_g$ plane is confirmed in the future, this would point towards the LQ representation $\tilde{\Phi}_2$. Similarly, one can correlate $\kappa_\gamma$ to $\kappa_{Z\gamma}$, see Fig.~\ref{Zgamma}, which clearly provides complementary distinguishing power, especially at the FCC-hh. E.g. an anti-correlations between $\kappa_\gamma$ to $\kappa_{Z\gamma}$ is not favored by either (single) Lagrangian parameter of coupling LQs to the Higgs.

\section{Conclusions}
\label{sec:Conclusions}

LQs are prime candidates to explain the flavor anomalies, i.e. the discrepancies between the SM predictions and experiment in $b\to c\tau\nu$ and $b\to s\ell^+\ell^-$ processes and in the anomalous magnetic moment of the muon. Therefore, it is interesting to study alternative observables which are sensitive to LQs and could therefore as well show deviations from the SM predictions. In this context, parameters sensitive to additional electroweak symmetry breaking effects provide a complementary window. In particular, LQ couplings to the SM Higgs generate loop effects, which contribute to the oblique parameters ($S$ and $T$) and to effective Higgs couplings, entering on-shell Higgs boson production ($gg\to h$) and decays ($h\to \gamma\gamma$, $h\to Z\gamma$). All these observables have in common that (at the one-loop level) they do not depend on the LQ couplings to fermions but rather only on LQ couplings to Higgses (tri-linear and quadratic ones). Therefore, one can test this sector of the Lagrangian independently of the fermion couplings entering flavor observables. 

Taking into account the most general set of Higgs-LQ interactions, including mixing among different LQ representations, we calculated the one-loop contributions to the oblique parameters $S$, $T$ and $U$. Using a perturbative expansion of the mixing matrices we were able to provide simple, analytic expressions for them. Similarly, we calculated the contributions to effective on-shell $hgg$, $h\gamma\gamma$ and $hZ\gamma$ couplings, expressing the corrections as simple analytic functions of the Lagrangian parameters.

In our phenomenological analysis we correlated the effects in the oblique corrections with each other, see Fig.~\ref{STcorrelations}, finding that the contribution to $T$ is positive definite and that $T$ is clearly more sensitive to LQs than $S$. Similarly, we correlated $hgg$ with $h\gamma\gamma$ in Fig.~\ref{kappacorrelations} and $h\gamma\gamma$ to $hZ\gamma$ in Fig.\ref{Zgamma}. In the future it would be very interesting to include the NLO QCD corrections, in the spirit of Refs.~\cite{Muhlleitner:2006wx,Bonciani:2007ex}, as these interesting correlations open the possibility of distinguishing different LQ representations, independently of their couplings to fermions, providing strong motivation for future colliders.  

\bigskip

{\it Acknowledgements} --- {We thank Michael Spira for useful discussions. The work of A.C. and D.M. is supported by a Professorship Grant (PP00P2\_176884) of the Swiss National Science Foundation. The work of F.S. is supported by the Swiss National Foundation grant 200020\_175449/1.}
\bigskip

\appendix
\section{Appendix}

\subsection{Loop Functions}
In this appendix we first present the loop functions, which are used in \eqref{eq:oblique_parameters} to write the results for the $S$ and $T$ parameters in a compact form
\begin{align*}
\mathcal{K}_{1}(y)&=-10\bigg(\frac{y^3+2y^2-19y+4}{(y-1)^4}-\frac{(4y^3-12y^2-6y+2)\log(y)}{(y-1)^5}\bigg)\,,\\
\mathcal{K}_{2}(y)&=\frac{10}{17}\bigg(\frac{-y^4+10y^3-45y^2-8y+8}{y(y-1)^4}+\frac{18(3y-1)\log(y)}{(y-1)^5}\bigg)\,,\\
\mathcal{K}_{3}(y)&=10\bigg(\frac{y^2+10y+1}{(y-1)^4}-\frac{6y(y+1)\log(y)}{(y-1)^5}\bigg)\,,\\
\mathcal{K}_{4}(y)&=2\bigg(\frac{(y+4)(y^2+10y+1)}{y(y-1)^4}-\frac{6(y+1)(y+4)\log(y)}{(y-1)^5}\bigg)\,,\\
\mathcal{K}_{5}(y)&=2\bigg(\frac{2y^2+5y-1}{(y-1)^3}-\frac{6y^2\log(y)}{(y-1)^4}\bigg)\,,\\
\mathcal{K}_6(y)&=2 \bigg( \frac{y^2-5y-2}{(y-1)^3y} + \frac{6}{(y-1)^4}\bigg)
\\
\mathcal{K}_{7}(x)&=\frac{3\big(x^2-1-2x\log(x)\big)}{(x-1)^3}\,,\\
\mathcal{K}_{8}(x,y)&=-3\bigg(\frac{4}{(x-1)^2(y - 1)} + \frac{8x}{(x - 1)(y - x)^2} - \frac{4}{(x - 1)^2(y-x)}\\ &+ 4x \log(x) \,\mathcal{K}_{10}(x,y) + 4y \log(y)\,\mathcal{K}_{10}(y,x)\bigg) \ , \\ 
\mathcal{K}_9(x,y)&=10\bigg(\frac{12x}{(1-x)^2(x-y)^2} - \frac{2x^2-7x-13}{2(x-1)^3(y-1)} - \frac{6(x+1)}{(x-1)^3(y-x)} \\ &- \frac{9(x-3)}{2(x-1)^2(y-1)^2}-\frac{3}{(x-1)(y-1)^3} \\
&+ 3x \log(x) \,\mathcal{K}_{11}(x,y) + 3y \log(y)\,\mathcal{K}_{11}(y,x)\bigg) \ , \\ 
\intertext{and}
\mathcal{K}_{10}(x,y) &= \frac{2x}{(x - 1)(y - x)^3} + \frac{x - 2}{(x - 1)^2(y - x)^2} \ , \\
\mathcal{K}_{11}(x,y) &= \frac{4x}{(x-1)^2(y-x)^3} - \frac{4}{(x-1)^3(y -x)^2} + 
\frac{x}{(x - 1)^4(y - x)} \ .
\end{align*}
In $h\to Z\gamma$ we used the following loop functions for the amplitude
\begin{align*}
\mathcal{F}_1(x)&=2 \left(\frac{x^3-6x^2+3x+6x \log(x) +2}{(x-1)^4}\right) \ , \\
\mathcal{F}_2(x)&=\frac{2}{3}\left( \frac{x^4-2x^3+9x^2-6x^2 \log(x) -10x + 2}{x(x-1)^4}\right) \ .
\end{align*}

\subsection{Expanded Matrices}
Next, we will give the expressions for the coupling matrices, expanded in terms of the vacuum expectation value $v$. We have the weak isospin matrices $T^{Q}$, which read in case of no LQ mixing
\begin{align}
\begin{aligned}
T^{-1/3}=
\begin{pmatrix}
0&0&0\\ 0&-\frac{1}{2}&0 \\ 0&0&0
\end{pmatrix}\,,
&&
T^{2/3}=
\begin{pmatrix}
-\frac{1}{2}&0&0\\ 0&\frac{1}{2}&0\\ 0&0&1
\end{pmatrix}\,,&&
T^{-4/3}=
\begin{pmatrix}
0&0\\ 0&-1
\end{pmatrix}\,,
&&
T^{5/3}=\frac{1}{2}\,,
\end{aligned}
\end{align}
using the basis defined in Eq. \eqref{EWeigenstates}. A unitary redefinition of the LQ fields in order to diagonalize the mass matrices in Eq. \eqref{eq:LQ_mixing_matrices} also affects the $T^{Q}$ matrices
\begin{align}
\tilde{T}^{Q}=W^{Q}T^{Q}W^{Q\dagger}\,.
\end{align}
Note that the LQ field redefinition has no impact the electromagnetic interaction, since the coupling matrix is proportional to the unit matrix and the $W^{Q}$ then cancel due to unitarity. If we use the perturbative diagonalization ansatz, we obtain 
\begin{align}
\begin{aligned}
\tilde{T}^{-1/3}&\!\approx\!
\begin{pmatrix}
\frac{-v^2|A_{\tilde 21}|^2}{2(m_{1}^2-\tilde{m}_{2}^2)^2} & \frac{vA_{\tilde 21}^{*}}{2(\tilde{m}_2^2 -m_{1}^2)} & \frac{v^2 A_{3\tilde 2}A_{\tilde 21}^{*}}{2(m_{1}^2-\tilde{m}_{2}^2)(\tilde{m}_{2}^2-m_{3}^2)}\\
\frac{vA_{\tilde 21}}{2(\tilde{m}_{2}^2-m_1^2)}& -\frac{1}{2}\!+\!\frac{v^2}{2}\Big(\!\frac{|A_{\tilde 21}|^2}{(m_{1}^2-\tilde{m}_{2}^2)^2}\!+\!\frac{|A_{3\tilde 2}|^2}{(\tilde{m}_{2}^2-m_{3}^2)^2}\!\Big) & \frac{v A_{3\tilde 2}}{2(\tilde{m}_{2}^2-m_3^2)}\\
\frac{v^2 A_{\tilde 21}A_{3\tilde 2}^{*}} {2(m_{1}^2-\tilde{m}_{2}^2)(\tilde{m}_{2}^2-m_{3}^2)}& \frac{vA_{3\tilde 2}^{*}}{2(\tilde{m}_{2}^2-m_3^2)}& \frac{-v^2|A_{3\tilde 2}|^2}{2(\tilde{m}_{2}^2-m_{3}^2)^2}
\end{pmatrix}\,,\\
\tilde{T}^{2/3}&\!\approx\!
\begin{pmatrix}
-\frac{1}{2} & \frac{v^2 Y_{\tilde 22}}{m_2^2-\tilde{m}_{2}^2} & 0\\
\frac{v^2 Y_{\tilde 22}^{*}}{m_2^2-\tilde{m}_{2}^2} & \frac{1}{2}\!+\!\frac{v^2|A_{3\tilde 2}|^2}{(\tilde{m}_{2}^2-m_{3}^2)^2} &\frac{vA_{3\tilde 2}}{\sqrt{2}(m_3^2-\tilde{m}_{2}^2)}\\
0 & \frac{v A_{3\tilde 2}^{*}}{\sqrt{2}(m_3^2-\tilde{m}_{2}^2)} & 1\!-\!\frac{v^2 |A_{3\tilde 2}|^2}{(\tilde{m}_{2}^2-m_{3}^2)^2}
\end{pmatrix}\,,\\
\tilde{T}^{-4/3}&\!\approx\!
\begin{pmatrix}
0 & \frac{\sqrt{2}v^2 Y_{3 \tilde 1}^{*}}{m_3^2-\tilde{m}_{1}^2}\\
\frac{\sqrt{2}v^2 Y_{3 \tilde 1}}{m_3^2-\tilde{m}_{1}^2} & -1
\end{pmatrix}\,,
\end{aligned}
\label{eq:T3_mixing}
\end{align}
valid up to $\mathcal{O}(v^2)$. $T^{5/3}$ is not affected, since the LQ with charge $Q=5/3$ does not mix.\\
There are also interaction matrices for the $ZZ\Phi^Q \Phi^Q$ vertex, which read in case of no LQ mixing
\begin{align}
\begin{aligned}
D^{-1/3}&= \begin{pmatrix}
\Big(\frac{s_w^2}{3}\Big)^{\!2} & 0 & 0 \\
0 & \Big(\frac{s_w^2}{3} -\frac{1}{2}\Big)^{\!2}  & 0 \\
0 & 0 & \Big(\frac{s_w^2}{3}\Big)^{\!2} 
\end{pmatrix} &&
D^{2/3}= \begin{pmatrix}
\Big(\frac{2 s_w^2}{3} + \frac{1}{2}\Big)^{\!2}  & 0 & 0 \\
0 & \Big(\frac{2 s_w^2}{3}-\frac{1}{2} \Big)^{\!2}  & 0 \\
0 & 0 & \Big( \frac{2 s_w^2}{3} -1\Big)^{\!2} 
\end{pmatrix} \\
D^{-4/3}& = \begin{pmatrix}
\left(\frac{4 s_w^2}{3} \right)^{\!2}  & 0 \\
0 & \Big( \frac{4 s_w^2}{3} -1 \Big)^{\!2} 
\end{pmatrix}
&& D^{5/3}=\Big(\frac{5s_w^2}{3}-\frac{1}{2}\Big)^{\!2} \,.
\end{aligned}
\end{align}
If we include the LQ mixing, we have
\begin{align}
\begin{aligned}
\tilde{D}^{-1/3}&\approx\frac{1}{12}
\begin{pmatrix}
\frac{4s_{w}^4}{3}-\frac{(4s_{w}^2-3)v^2|A_{\tilde{2}1}|^2}{(m_{1}^2-\tilde{m}_{2}^2)^2} & \frac{(4s_{w}^2-3)v A_{\tilde{2}1}^{*}}{\tilde{m}_{2}^2-m_{1}^2} & \frac{(4s_{w}^2-3)v^2 A_{\tilde{2}1}^{*}A_{3\tilde{2}}}{(m_{3}^2-\tilde{m}_{2}^2)(\tilde{m}_{2}^2-m_{1}^2)}\\
\frac{(4s_{w}^2-3)vA_{\tilde{2}1}}{\tilde{m}_{2}^2-m_{1}^2} & \tilde{d}_{22} & \frac{(4s_{w}^2-3)v A_{3\tilde{2}}}{\tilde{m}_{2}^2-m_{3}^2}\\
\frac{(4s_{w}^2-3)v^2 A_{3\tilde{2}}^{*}A_{\tilde{2}1}}{(m_{1}^2-\tilde{m}_{2}^2)(\tilde{m}_{2}^2-m_{3}^2)} & \frac{(4 s_{w}^2-3)vA_{3\tilde{2}}^{*}}{\tilde{m}_{2}^2-m_{3}^2} & \frac{4s_{w}^4}{3}-\frac{(4s_{w}^2-3)v^2|A_{3\tilde{2}}|^2}{(m_{3}^2-\tilde{m}_{2}^2)^2}
\end{pmatrix}\,,\\
\text{with}&\quad \tilde{d}_{22}=\frac{(3-2s_{w}^2)^2}{3}+\frac{(4s_{w}^2-3)v^2|A_{\tilde{2}1}|^2}{(m_{1}^2-\tilde{m}_{2}^2)^2}+\frac{(4s_{w}^2-3)v^2|A_{3\tilde{2}}|^2}{(m_{3}^2-\tilde{m}_{2}^2)^2}\,,\\
\tilde{D}^{2/3}&\approx\frac{1}{12}
\begin{pmatrix}
\frac{(4s_{w}^2+3)^2}{3} & \frac{16 s_{w}^2v^2 Y_{\tilde{2}2}}{\tilde{m}_{2}^2-m_{2}^2} & 0\\
\frac{16s_{w}^2 v^2 Y_{\tilde{2}2}^{*}}{\tilde{m}_{2}^2-m_{2}^2} & \frac{(4s_{w}^2-3)^2}{3}-\frac{2(8s_{w}^2-9)v^2|A_{3\tilde{2}}|^2}{(m_{3}^2-\tilde{m}_{2}^2)^2} & \frac{\sqrt{2}(8s_{w}^2-9)vA_{3\tilde{2}}}{\tilde{m}_{2}^2-m_{3}^2}\\
0 & \frac{\sqrt{2}(8s_{w}^2-9)v A_{3\tilde{2}}^{*}}{\tilde{m}_{2}^2-m_3^2}& \frac{4(3-2s_{w}^2)^2}{3}+\frac{2(8s_{w}^2-9)v^2|A_{3\tilde{2}}|^2}{(m_{3}^2-\tilde{m}_{2}^2)^2}
\end{pmatrix}\,,\\
\tilde{D}^{-4/3}&\approx\frac{1}{3}
\begin{pmatrix}
\frac{16s_{w}^4}{3} & \frac{\sqrt{2}(8s_{w}^2-3)v^2Y_{31}^{*}}{m_{3}^2-\tilde{m}_{1}^2}\\ \frac{\sqrt{2}(8s_{w}^2-3)v^2Y_{31}}{m_{3}^2-\tilde{m}_{1}^2} &\frac{(3-4s_{w}^2)^2}{3}
\end{pmatrix}\,.
\end{aligned}
\end{align}
Analogously to the $Z$ boson, different LQ generations mix under $W$ interactions. Without LQ mixing, the interactions with the $W$ boson can be written in terms of the following matrices
\begin{align}
B^{1}=
\begin{pmatrix}
0 & 0 & 0\\
0 & 0 & \sqrt{2}
\end{pmatrix}\,,&&
B^{2}=
\begin{pmatrix}
0 & 0 & 0 \\
0 & 1 & 0 \\
0 & 0 & -\sqrt{2}
\end{pmatrix}\,,&&
B^{3}=
\begin{pmatrix}
1 & 0 & 0
\end{pmatrix}\,,
\end{align}
arranging the LQ in their charge eigenstates according to Eq. \eqref{EWeigenstates}. $B^{1}$ describes the interaction of LQs with electric charges $Q=-4/3$ and $Q=-1/3$, $B^{2}$ the ones with $Q=-1/3$ and $Q=2/3$, $B^{3}$ with $Q=5/3$ and $Q=2/3$. If we include LQ mixing, the matrices expanded up to $\mathcal{O}(v^2)$, then read
\begin{align}
\tilde{B}^{1}&\!\approx\!
\begin{pmatrix}
0 & 0 & \frac{2v^2 Y_{3 \tilde 1}^{*}}{\tilde{m}_{1}-m_{3}^2}\\
\frac{\sqrt{2}v^2}{m_{1}^2-m_{3}^2}\Big(\!\frac{A_{\tilde 21}A_{3\tilde 2}^{*}}{m_1^2-\tilde{m}_{2}^2}\!+\!Y_{31}^{*}\!\Big) & \frac{\sqrt{2}vA_{3\tilde 2}^{*}}{\tilde{m}_{2}^2-m_{3}^2} & \sqrt{2}\!-\!\frac{v^2|A_{3\tilde 2}|^2}{\sqrt{2}(\tilde{m}_{2}^2-m_{3}^2)^2}
\end{pmatrix}\,,\nonumber\\
\tilde{B}^{2}&\!\approx\!
\begin{pmatrix}
0 & \frac{vA_{\tilde 21}^{*}}{m_{1}^2-\tilde{m}_{2}^2} & \frac{\sqrt{2}v^2}{m_{1}^2-m_{3}^2}\Big(\!\frac{A_{3\tilde 2}A_{\tilde 21}^{*}}{m_{3}^2-\tilde{m}_{2}^2}\!-\!Y_{31}\!\Big)\\
\frac{v^2 Y_{\tilde 22}^{*}}{m_{2}^2-\tilde{m}_{2}^2} & 1\!-\!\frac{v^2}{2}\Big(\!\frac{|A_{\tilde 21}|^2}{(m_{1}^2-\tilde{m}_{2}^2)^2}\!-\!\frac{|A_{3\tilde 2}|^2}{(\tilde{m}_{2}^2-m_{3}^2)^2}\!\Big) & 0\\
0 & \frac{vA_{3\tilde 2}^{*}}{\tilde{m}_{2}^2-m_{3}^2} & -\sqrt{2}\!-\!\frac{v^2|A_{3\tilde 2}|^2}{\sqrt{2}(m_{3}^2-\tilde{m}_{2}^2)^2}
\end{pmatrix}\,,\nonumber\\
\tilde{B}^{3}&\!\approx\!
\begin{pmatrix}
1 & \frac{-v^2 Y_{\tilde 22}^{*}}{m_{2}^2-\tilde{m}_{2}^2} & 0
\end{pmatrix}\,.
\label{eq:BW_mixing}
\end{align}
We also have interaction matrices for the $W^{+}W^{-}\Phi^Q \Phi^Q$ vertex. Without mixing, they read
\begin{align}
F^{-1/3} = \begin{pmatrix}
0 & 0 & 0 \\ 0 & \frac{1}{2} & 0 \\ 0 & 0 & 2
\end{pmatrix} &&
F^{2/3} = \begin{pmatrix}
\frac{1}{2} & 0 & 0 \\ 0 & \frac{1}{2} & 0 \\ 0 & 0 & 1
\end{pmatrix} &&
F^{-4/3} = \begin{pmatrix}
0 & 0 \\ 0 & 1
\end{pmatrix} \ ,
&&
F^{5/3}=\frac{1}{2}\,.
\end{align}
If we include mixing and expand up to order $\mathcal{O}(v^2)$, we obtain
\begin{align}
\begin{aligned}
\tilde{F}^{-1/3} &\!\approx\! \begin{pmatrix}
\frac{v^2|A_{\tilde 21}|^2}{2 (m_1^2 - \tilde{m}_2^2)^2} & \frac{v A_{\tilde 21}^*}{2(m_1^2-\tilde{m}_2^2)} &  \tilde{f}_{13}
\\
\frac{v A_{\tilde 21}}{2(m_1^2-\tilde{m}_2^2)} &  \frac{1}{2} + \frac{v^2}{2} \left( \frac{3 |A_{3\tilde 2}|^2}{(\tilde{m}_2^2-m_3^2)^2} - \frac{|A_{\tilde 21}|^2}{(\tilde{m}_2^2-m_1^2)^2} \right) & \frac{3 v A_{3\tilde 2}}{2(\tilde{m}_2^2-m_3^2)} \,,
\\
\tilde{f}_{13}^{*} & \frac{3 v A_{3\tilde 2}^*}{2(\tilde{m}_2^2-m_3^2)} & 2 - \frac{3 v^2 |A_{3\tilde 2}|^2}{2 (\tilde{m}_2^2-m_3^2)^2}\,,
\end{pmatrix}\,,\\
\text{with}&\quad \tilde{f}_{13}=\frac{2 v^2 Y_{31}}{m_1^2-m_3^2} - \frac{v^2 A_{3\tilde 2} A_{\tilde 21}^* (m_1^2-4\tilde{m}_2^2+3m_3^2)}{2(m_1^2-m_3^2)(m_1^2-\tilde{m}_2^2)(\tilde{m}_2^2-m_3^2)}\\
\tilde{F}^{2/3}&\!\approx\!
\begin{pmatrix}
\frac{1}{2} & 0 & 0 
\\
0 & \frac{1}{2} + \frac{v^2 |A_{3\tilde 2}|^2}{(\tilde{m}_2^2-m_3^2)^2} & \frac{v A_{3\tilde 2}}{\sqrt{2} (-m_3^2\tilde{m}_2^2)}
\\
0 & \frac{v A_{3\tilde 2}^*}{\sqrt{2} (m_3^2-\tilde{m}_2^2)}  & 1 - \frac{v^2 |A_{3\tilde 2}|^2}{(\tilde{m}_2^2-m_3^2)^2}
\end{pmatrix}\,,
\\
\tilde{F}^{-4/3} &\!\approx\!
\begin{pmatrix}
0 & \frac{\sqrt{2} v^2 Y_{3 \tilde 1}^*}{m_1^2-m_3^2} \\ 
\frac{\sqrt{2} v^2 Y_{3 \tilde 1}}{m_1^2-m_3^2} & 1
\end{pmatrix}\,.
\end{aligned}
\end{align}
Finally we show the Higgs coupling matrices in \eqref{eq:higgs_couplings} up to $\mathcal{O}(v)$
\begin{align}
\begin{aligned}
\tilde{\Gamma}^{-1/3}&\approx\frac{1}{\sqrt{2}}
\begin{pmatrix}
2v\big(Y_{1}+\frac{|A_{\tilde{2}1}|^2}{m_{1}^2-\tilde{m}_{2}^2}\big) & A_{\tilde{2}1}^{*} & v\Big(2Y_{31}-\frac{A_{3\tilde{2}}A_{\tilde{2}1}^{*}(m_{1}^2+m_{3}^2-2\tilde{m}_{2}^2)}{(m_{1}^2-\tilde{m}_{2}^2)(\tilde{m}_{2}^2-m_{3}^2)}\Big)\\
A_{\tilde{2}1} & \tilde{\Gamma}_{22}^{-1/3} & A_{3\tilde{2}}\\
v\Big(2Y_{31}^{*}-\frac{A_{\tilde{2}1}A_{3\tilde{2}}^{*}(m_{1}^2+m_{3}^2-2\tilde{m}_{2}^2)}{(m_{1}^2-\tilde{m}_{2}^2)(\tilde{m}_{2}^2-m_{3}^2)}\Big)&A_{3\tilde{2}}^{*} & 2v\Big(Y_{3}+\frac{|A_{3\tilde{2}}|^2}{m_{3}-\tilde{m}_{2}^2}\Big)
\end{pmatrix}\,,\\
\text{with}&\quad \tilde{\Gamma}_{22}^{-1/3}=2v\Big(Y_{\tilde{2}}-\frac{|A_{\tilde{2}1}|^2}{m_{1}^2-\tilde{m}_{2}^2}-\frac{|A_{3\tilde{2}}|^2}{m_{3}^2-\tilde{m}_{2}^2}\Big)\,,\\
\tilde{\Gamma}^{2/3}&\approx \frac{1}{\sqrt{2}}
\begin{pmatrix}
2v Y_{2} & 2v Y_{\tilde{2}2} & 0\\
2vY_{\tilde{2}2}^{*} & 2v\Big(Y_{\tilde{2}}+Y_{\tilde{2}\tilde{2}}-\frac{2|A_{3\tilde{2}}|^2}{m_{3}^2-\tilde{m}_{2}^2}\Big) & -\sqrt{2}A_{3\tilde{2}}\\
0 & -\sqrt{2}A_{3\tilde{2}}^{*} & 2v\Big(Y_{3}+\frac{2|A_{3\tilde{2}}|^2}{m_{3}^2-\tilde{m}_{2}^2}\Big)
\end{pmatrix}\,,\\
\tilde{\Gamma}^{-4/3}&\approx \Gamma^{-4/3}\,.
\end{aligned}
\end{align}
and
\begin{align}
\begin{aligned}
\tilde{\Lambda}^{-1/3}&\approx \frac{1}{2}
\begin{pmatrix}
Y_{1}& v\Big(\frac{Y_{31}A_{3\tilde{2}}^{*}}{\tilde{m}_{2}^2-m_{3}^2}+\frac{(Y_{\tilde{2}}-Y_{1})A_{\tilde{2}1}^{*}}{m_{1}^2-\tilde{m}_{2}^2}\Big) & Y_{31} \\
v\Big(\frac{Y_{31}^{*}A_{3\tilde{2}}}{\tilde{m}_{2}^2-m_{3}^2}+\frac{(Y_{\tilde{2}}-Y_{1})A_{\tilde{2}1}}{m_{1}^2-\tilde{m}_{2}^2}\Big) & Y_{\tilde{2}} & v\Big(\frac{Y_{31}A_{\tilde{2}1}}{\tilde{m}_{2}^2-m_{1}^2}+\frac{(Y_{\tilde{2}}-Y_{3})A_{3\tilde{2}}}{m_{3}^2-\tilde{m}_{2}^2}\Big)\\
Y_{31}^{*} & v\Big(\frac{Y_{31}^{*}A_{\tilde{2}1}^{*}}{\tilde{m}_{2}^{2}-m_{1}^2}+\frac{(Y_{\tilde{2}}-Y_{3})A_{3\tilde{2}}^{*}}{m_{3}^2-\tilde{m}_{2}^2}\Big) & Y_{3}
\end{pmatrix}\,,\\
\tilde{\Lambda}^{2/3}&\approx \frac{1}{2}
\begin{pmatrix}
Y_{2} & Y_{\tilde{2}2} & \frac{\sqrt{2}vY_{\tilde{2}2}A_{3\tilde{2}}}{\tilde{m}_{2}^2-m_{3}^2} \\
Y_{\tilde{2}2}^{*} & Y_{\tilde{2}}+Y_{\tilde{2}\tilde{2}} & \frac{\sqrt{2}v(Y_{3}-Y_{\tilde{2}}-Y_{\tilde{2}\tilde{2}})A_{3\tilde{2}}}{m_{3}^2-\tilde{m}_{2}^2}\\
\frac{\sqrt{2}vY_{\tilde{2}2}^{*}A_{3\tilde{2}}^{*}}{\tilde{m}_{2}^2-m_{3}^2} & \frac{\sqrt{2}v(Y_{3}-Y_{\tilde{2}}-Y_{\tilde{2}\tilde{2}})A_{3\tilde{2}}^{*}}{m_{3}^2-\tilde{m}_{2}^2} & Y_{3}
\end{pmatrix}\,,\\
\tilde{\Lambda}^{-4/3}&\approx\Lambda^{-4/3}\,.
\end{aligned}
\end{align}

\begin{boldmath}
\subsection{Exact Results for the Vacuum Polarization Functions}
\end{boldmath}
In this section we give the $q^2$-expanded results for the vacuum polarization functions, with the LQ masses and couplings kept unexpanded
\begin{align}
\begin{aligned}
\Pi_{\gamma\gamma}(q^2)&=-\sum_Q\frac{N_{c}\,e^2 Q^{2}}{48\pi^2}q^2\left(\frac{1}{\varepsilon}+\log\!\left(\frac{\mu^2}{{(M_{a}^Q)}^2}\right)+\frac{q^2}{10 {(M_{a}^Q)}^2}\right)\,,\\
\Pi_{Z\gamma}(q^2)&=-\sum_Q\frac{N_{c}\,g_{2}e\left(\tilde{T}^{Q}_{aa}-s_{w}^2 Q\right)Q}{48\pi^2 c_{w}}q^2\left(\frac{1}{\varepsilon}+\log\!\left(\frac{\mu^2}{{(M_{a}^Q)}^2}\right)+\frac{q^2}{10 {(M_{a}^Q)}^2}\right)\,,\\
\Pi_{ZZ}(q^2)&=\sum_Q\frac{N_{c}}{16\pi^2}\left(\frac{g_{2}\Gamma_{aa}^{Q}}{c_{w}}\frac{m_{Z}}{m_{h}^2}-\frac{2g_{2}^2\tilde{D}_{aa}^{Q}}{c_{w}^2}\right){(M_{a}^Q)}^2\left(\frac{1}{\varepsilon}+\log\!\left(\frac{\mu^2}{{(M_{a}^Q)}^2}\right)+1\right)\\
&-\sum_Q\frac{N_{c}\,g_{2}^2\left(\tilde{T}^{Q}_{ab}-s_{w}^2 Q\delta_{ab}\right)\left(\tilde{T}^{Q}_{ba}-s_{w}^2 Q\delta_{ba}\right)}{32\pi^2c_{w}^2}\times F\!\left({(M_{a}^Q)}^2,\, {(M_{b}^Q)}^2\right)\,,\\
\Pi_{WW}(q^2)&=\sum_Q\frac{N_{c}}{16\pi^2}\left(g_{2}\Gamma_{aa}^{Q}\frac{m_{W}}{m_{h}^2}-g_{2}^2\tilde{F}^{Q}_{aa}\right){(M_{a}^Q)}^2\left(\frac{1}{\varepsilon}+\log\!\left(\frac{\mu^2}{{(M_{a}^Q)}^2}\right)+1\right)\\
&-\sum_{a(b)=1}^{2(3)}\frac{N_{c}\,g_{2}^2\left|\tilde{B}^{1}_{ab}\right|^2}{64\pi^2}{(M_{a}^{-4/3})}^2F\!\left({(M_{a}^{-4/3})}^2,\, {(M_{b}^{-1/3})}^2\right)
\\
&-\sum_{a,b=1}^3\frac{N_{c}\,g_{2}^2\left|\tilde{B}^{2}_{ab}\right|^2}{64\pi^2}{(M_{a}^{-1/3})}^2F\!\left({(M_{a}^{-1/3})}^2,\, {(M_{b}^{2/3})}^2\right)
\\
&-\sum_{b=1}^3\frac{N_{c}\,g_{2}^2\left|\tilde{B}^{3}_{b}\right|^2}{64\pi^2}{(M^{5/3})}^2F\!\left({(M^{5/3})}^2,\, {(M_b^{2/3})}^2\right)\,,
\end{aligned}
\end{align}
where $Q=\{-1/3, 2/3, -4/3, 5/3\}$ with $a$ and $b$ running from 1 to 3, 3, 2, and 1, respectively. Here we defined
\begin{equation}
F\left({(M_{a}^Q)}^2,\, {(M_{b}^Q)}^2\right)={(M_{a}^Q)}^2 \left(f_{0}+\frac{q^2}{{(M_{a}^Q)}^2}\,f_{1}+\left(\frac{q^2}{{(M_{a}^Q)}^2}\right)^{\!2}f_{2}\right)\,,
\end{equation}
with
\begin{align}
\begin{aligned}
f_{0}&=-2(x^Q_{ba}+1)\bigg(\frac{1}{\varepsilon}+\log\!\Big(\frac{\mu^2}{M_{a}^2}\Big)+\frac{3}{2}\bigg)+\frac{2(x^Q_{ba})^2\log(x^Q_{ba})}{x^Q_{ba}-1}\,,\\
f_{1}&=\frac{2}{3}\bigg(\frac{1}{\varepsilon}+\log\!\Big(\frac{\mu^2}{M_{a}^2}\Big)\bigg)-\frac{5-27x^Q_{ba}+27(x^Q_{ba})^2-5(x^Q_{ba})^3-6(3-x^Q_{ba})(x^Q_{ba})^2\log(x^Q_{ba})}{9(x^Q_{ba}-1)^3}\,,\\
f_{2}&=\frac{-1+8x^Q_{ba}-8(x^Q_{ba})^3+(x^Q_{ba})^4+12(x^Q_{ba})^2\log(x^Q_{ba})}{6(x^Q_{ba}-1)^5}\,,
\end{aligned}
\end{align}
where again
\begin{equation*}
x^Q_{ba}=\frac{{(M_{b}^Q)}^2}{{(M_{a}^Q)}^2} \ .
\end{equation*}

\subsection{Leading Order SM Amplitudes in Higgs Decays}
The SM amplitudes for the $h\gamma\gamma$, $hgg$ and $h Z\gamma$ couplings in Eq. \eqref{eq:SM_amplitude_Higgs} read
\begin{align}
\begin{aligned}
A_{1}(x)&=\frac{-(2x^2+3x+3(2x-1)f(x))}{x^2}\,,\\
A_{1/2}(x)&=\frac{2(x+(x-1)f(x))}{x^2}\\
C_{1}(x,y)&=4\big(3-\frac{s_{w}^2}{c_{w}^2}\big)I_{2}(x,y)+\Big(\big(1+\frac{2}{x}\big)\frac{s_{w}^2}{c_{w}^2}-\big(5+\frac{2}{x}\big)\Big)I_{1}(x,y)\,,\\
C_{1/2}(x,y)&=I_{1}(x,y)-I_{2}(x,y)\,,
\end{aligned}
\end{align}
with
\begin{align}
\begin{aligned}
f(x)&=\arcsin^2(\sqrt{x})\,,\\
g(x)&=\sqrt{\frac{1}{x}-1}\arcsin(\sqrt{x})\,,\\
I_{1}(x,y)&=\frac{xy}{2(x-y)}+\frac{x^2 y^2\big(f(x^{-1})-f(y^{-1})\big)}{2(x-y)^2}+\frac{x^2 y\big(g(x^{-1})-g(y^{-1})\big)}{(x-y)^2}\,,\\
I_{2}(x,y)&=\frac{-xy\big(f(x^{-1})-f(y^{-1})\big)}{2(x-y)}\,.
\end{aligned}
\end{align}

\newpage

\bibliographystyle{JHEP}
\bibliography{BIB}

\end{document}